\documentclass[12pt,letterpaper]{article}
\usepackage[left=2cm,right=2cm,top=2.5cm,bottom=2.5cm]{geometry}
\usepackage{enumitem}
\usepackage{lastpage}
\usepackage{eurosym}
\usepackage{hyperref}
\usepackage{fancyhdr}

\usepackage{algorithmicx}
\usepackage{algorithm}
\usepackage{algpseudocode}

\usepackage{color}

\usepackage[utf8]{inputenc}
\usepackage[english]{babel}
\usepackage{amsmath,mdframed}
\usepackage{mathtools}
\usepackage{amsfonts}
\usepackage{amssymb}
\usepackage{multirow}
\usepackage[dvipsnames]{xcolor}
\usepackage{tabularx}
\usepackage{setspace}

\usepackage{bm}
\usepackage{float}
\usepackage{graphicx}
\usepackage{caption}
\usepackage{subcaption}
\usepackage{comment}
\usepackage{array}
\usepackage{booktabs}
\newcolumntype{x}[1]{>{\centering}p{#1}}
\usepackage{natbib}
\usepackage{rotating}
\usepackage{authblk}

\renewcommand{\vec}[1]{\boldsymbol{\mathbf{#1}}}

\newtheorem{theorem}{Theorem}
\newtheorem{corollary}{Corollary}
\newtheorem{remark}{Remark}

\title{Penalized G-estimation for effect modifier selection in a structural nested mean model for repeated outcomes}
\author[1,*]{Ajmery Jaman}
\author[2]{Guanbo Wang}
\author[3]{Ashkan Ertefaie}
\author[4,6]{Michèle Bally}
\author[5]{Renée Lévesque}
\author[1]{Robert W.\ Platt}
\author[6,1]{Mireille E.\ Schnitzer}

\affil[1]{{\footnotesize Department of Epidemiology and Biostatistics, McGill University, Montreal, Canada}}
\affil[2]{{\footnotesize CAUSALab, Department of Epidemiology, Harvard T.H. Chan Schoole of Public Health, Boston, USA}}
\affil[3]{{\footnotesize Department of Biostatistics, Epidemiology and Informatics, University of Pennsylvania, Philadelphia, USA}}
\affil[4]{{\footnotesize Department of Pharmacy, Centre Hospital of University of Montreal, Montreal, Canada}}
\affil[5]{{\footnotesize Department of Medicine, University of Montreal, Montreal, Canada}}
\affil[6]{{\footnotesize Faculty of Pharmacy, University of Montreal, Montreal, Canada}}
\affil[*]{{\footnotesize Corresponding author: Ajmery Jaman, Email: ajmery.jaman@mail.mcgill.ca}}
\date{}

\begin{document}
\maketitle
\begin{abstract}
\noindent
Effect modification occurs when the impact of the treatment on an outcome varies based on the levels of other covariates known as effect modifiers. Modeling these effect differences is important for etiological goals and for purposes of optimizing treatment. Structural nested mean models (SNMMs) are useful causal models for estimating the potentially heterogeneous effect of a time-varying exposure on the mean of an outcome in the presence of time-varying confounding. A data-adaptive selection approach is necessary if the effect modifiers are unknown \textit{a priori} and need to be identified. Although variable selection techniques are available for estimating the conditional average treatment effects using marginal structural models or for developing optimal dynamic treatment regimens, all of these methods consider a single end-of-follow-up outcome. In the context of an SNMM for repeated outcomes, we propose a doubly robust penalized G-estimator for the causal effect of a time-varying exposure with a simultaneous selection of effect modifiers and prove the oracle property of our estimator. We conduct a simulation study for the evaluation of its performance in finite samples and verification of its double-robustness property. Our work is motivated by the study of hemodiafiltration for treating patients with end-stage renal disease at the Centre Hospitalier de l'Université de Montréal. We apply the proposed method to investigate the effect heterogeneity of dialysis facility on the repeated session-specific hemodiafiltration outcomes.\\

\end{abstract}
\setstretch{1}

\section{Introduction}

When a statistical goal is to estimate the causal effect of a treatment or an exposure on an outcome, effect modification or heterogeneity in the treatment effect may be of interest. Effect modification occurs when the effect of a treatment on an outcome differs according to the values of some other variables, usually pre-treatment covariates; such variables are called effect modifiers (EMs) \citep{vanderweele2007four}. Modeling these effect differences is important for etiological goals and for purposes of optimizing treatment. Analyzing effect modification has gained attention in recent years because of the popularization of precision medicine \citep{ashley2015precision} and dynamic treatment regimes  \citep{murphy2003optimal, chakraborty2013statistical,chakraborty2014dynamic,wallace2015doubly}, which deal with personalizing treatments by incorporating patient-level information to optimize expected outcomes. Structural nested mean models (SNMMs) are useful causal models for estimating the potentially heterogeneous effect of a time-varying exposure on the mean of an outcome in the presence of time-varying confounding \citep{robins1989analysis, robins1997causal, robins2007invited}.

Our methodological development is motivated by an application in nephrology regarding patients with end-stage renal disease, which is the final, permanent stage of chronic kidney disease. Patients with end-stage renal failure must undergo a kidney transplant or regular dialysis to survive for more than a few weeks because the kidneys no longer function properly. 
Hemodiafiltration (HDF) is a dialysis technique that is used for cleaning the blood from waste and excess fluid. HDF combines two processes \citep{ronco2007hemodiafiltration}: diffusion (where solute molecules passively move from the blood to the dialysis fluid) and convection (where larger molecules are cleared from the blood). Hemodiafiltration is routinely used for patients with end-stage renal disease treated at the University of Montreal Hospital Centre (CHUM) outpatient dialysis clinic and its affiliated ambulatory dialysis center (CED). 
With hemodiafiltration, a large volume of plasma water is ultrafiltered, and this requires the administration of a substitution fluid back to the patient in order to maintain fluid balance. Convection volume is calculated as the sum of the substitution volume and the ultrafiltration volume \citep{marcelli2015high}.
The effectiveness of hemodiafiltration is indicated by the convection volume attained during each session. The CONvective TRAnsport STudy \citep{grooteman2012effect} randomized controlled trial and a meta-analysis of individual patient-level data from randomized controlled trials showed that hemodiafiltration, compared to hemodialysis, reduced the risk of all-cause mortality by approximately 22\%, and of cardiovascular disease mortality by 31\%, over a median follow-up of 2.5 years in the hemodiafiltration treatment subgroup achieving high convection volumes \citep{peters2016haemodiafiltration}. These results suggest that one should aim for a convection volume of at least 24 liters per session \citep{chapdelaine2015optimization}. 
On average, convection volumes recorded at the outpatient dialysis clinic located at the CHUM are lower than those at CED. This has triggered an interest in investigating if there is any effect of the dialysis facility (CHUM vs.\ CED) on the convection volume or for which patients such an effect exists. Available data from hospital records include sociodemographic information, diagnoses, medications, blood test results, and dialysis treatment parameters of all dialysis sessions per patient in the study time frame (March 1st, 2017 to December 1st, 2021). Convection volume outcomes were measured at the end of each successful dialysis session, enabling us to investigate the average effect of the dialysis facility on the session-specific mean convection volume. Since it is currently unknown which measured variables could modify an effect of the dialysis facility on the hemodiafiltration outcome, it is important to develop a data-adaptive approach to selecting effect modifiers in this context.

Variable selection techniques are available in estimation of the average causal effect  \citep{shortreed2017outcome, koch2018covariate,tang2023ultra}, the conditional average treatment effects using marginal structural models  \citep{bahamyirou2022doubly}, and in developing optimal dynamic treatment regimens  \citep{gunter2011variable,shi2018high,wallace2019model,bian2021variable,jones2022valid,bian2023variable,moodie2023variable}. All of these methods were largely developed for dynamic or non-dynamic treatment interventions, some in longitudinal settings, where the outcome is measured at a single point in time. Under a traditional semiparametric regression approach (targeting prediction), \cite{johnson2008penalized} proposed penalized estimating equations for variable selection with a univariate outcome, whereas \cite{wang2012penalized} extended this work to high-dimensional longitudinal data proposing penalized generalized estimating equations. \cite{boruvka2018assessing} discussed assessments of causal effect moderation (modification) in mobile health interventions, where treatment, response and potential moderators are all time-varying. A recent study proposed sequential knockoffs for variable selection in Markov decision process framework to address reinforcement learning problems \citep{ma2023sequential}. However, to the best of our knowledge, no method exists that conducts simultaneous variable selection and causal effect estimation with longitudinal data and repeated outcomes.  Motivated by our application, we seek to develop a doubly robust estimator for the causal effect of an exposure with simultaneous selection of effect modifiers in the SNMM for repeated outcomes. Our estimator facilitates sparse modeling, which is important for developing a tractable model when covariate dimensionality is high and improves precision in estimation by eliminating spurious effect modifiers.

This paper is organized as follows. In Section~\ref{method}, we introduce the notation, describe the model and assumptions, present methodological details, and provide theoretical results for the asymptotic properties of the proposed estimator.  We present a simulation study to evaluate the performance of our estimator in finite samples in Section~\ref{simulation}, and describe the application of the proposed method to the hemodiafiltration data in Section~\ref{application}. Finally, we provide a discussion in Section~\ref{discussion}.

\section{Methodology} \label{method}

\subsection{Notation and model with assumptions} \label{notation}
\begin{figure}
 \centerline{\includegraphics[scale=0.5]{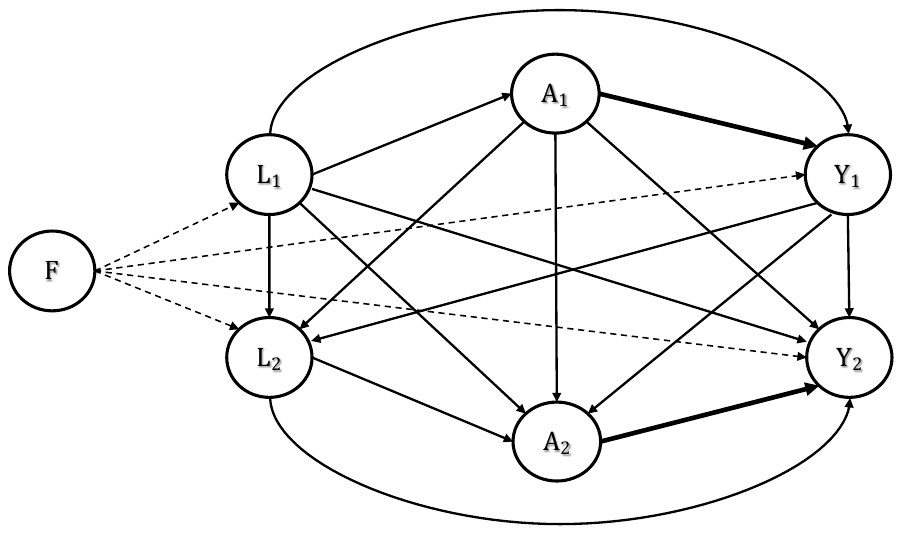}}
\caption{An example causal DAG representing the relationship among confounders, exposures, and outcomes in our time-varying setup.}
\label{fig1}
\end{figure}

Suppose we have data from $n$ patients and all  patients have measurements from $J$ sequential hemodiafiltration sessions. At each session, the information of the outcome, the treatment received, and pre-session covariates are recorded. For  patient $i$ at session $j$, denote the observed continuous outcome by $Y_{ij}$, the (binary) treatment received by $A_{ij}$, and the vector of covariates by $\vec L_{ij}, \forall i=1,\dots,n, j=1,\dots,J$. Let $\vec H_{ij}$ 
represent the histories of covariates $\overline{\vec L}_{ij}=\{\vec L_{i1},\ldots \vec L_{ij}\}$, past exposures $\overline{A}_{i(j-1)}=\{A_{i1},\ldots,A_{i(j-1)}\}$ and past outcomes $\overline{Y}_{i(j-1)}=\{Y_{i1},\ldots,Y_{i(j-1)}\}$. Throughout this paper, we use the potential outcomes framework \citep{robins1989analysis}. Define
$Y_{ij}(\overline{a}_{j})$ as the counterfactual outcome that would have been observed at occasion $j$ for patient $i$ if the treatment history $\overline{A}_{ij}=\{A_{i1},\ldots,A_{ij}\}$ were set counterfactually to $\overline{a}_j=\{a_1,\ldots,a_j\}$.
To model the proximal (short-term) effects of the exposure, a linear SNMM can be defined as \citep{robins1989analysis,vansteelandt2014structural}
\begin{align} \label{snmm.proximal}
\text{E}\{Y_{ij}(\overline{a}_{j-1}, a_j)-Y_{ij}(\overline{a}_{j-1}, 0)|\vec H_{ij}=\vec h_{ij},A_{ij}=a_{j}\}= \gamma_j^*(a_{j},\vec h_{ij};\vec\psi)
\end{align}
for each measurement occasion $j=1,\ldots,J$, where $\gamma_j^*(a_{j},\vec h_{ij};\vec\psi)$, known as ``treatment blip", is a scalar-valued function smooth in $\vec\psi$; $\vec h_{ij}$ represent the realized values for $\vec H_{ij}$; and $\vec\psi=(\psi_0,\psi_1,\ldots,\psi_{K-1})^\top$ is a K-dimensional vector of parameters. The difference in (\ref{snmm.proximal}) 
represents the effect of treatment $a_j$ versus the reference treatment 0 on the outcome at occasion $j$, given the history up to occasion $j$.  Our goal is thus to estimate the parameters $\vec\psi$ using the observed data. The following assumptions are required to identify $\vec\psi$ from the observed data \citep{robins1989analysis,vansteelandt2014structural,he2015structural}.
\begin{itemize}
    \item Consistency: The observed outcome is equal to the potential outcome at occasion $j$, for $j=1,\dots,J$, if the observed treatment history matches the counterfactual history at occasion $j$, i.e., $Y_{ij}(\overline{a}_j) = Y_{ij}$, if $\overline{A}_{ij}=\overline{a}_{j}$;
    \item Sequential ignorability/conditional exchangeability: The potential outcome $Y_{ij}(\overline{a}_{j-1},0)$ is independent of $A_{ij}$ conditional on $\vec H_{ij}$, for $j=1,\dots,J$.
    \item Positivity: If the joint density of $\vec H_{ij}$ at $\{\vec h_{ij}\}$ is greater than zero, then $\Pr(A_{ij}=a_j|\vec H_{ij}=\vec h_{ij}) > 0$ for all $a_j$, $j=1,\dots,J$.
\end{itemize}
To visualize this time-varying setup, the data generating process for the first two measurement occasions is shown in the directed acyclic graph (DAG) in Figure~\ref{fig1}. In the figure, $F$ represents the unmeasured variables which are not confounders. The bold arrows in the figure show the proximal effect of the treatment that we want to estimate.

\subsection{G-estimation with repeated outcomes} \label{sec.G.estimation}
G-estimation is used for estimating the parameters in SNMMs \citep{robins2008estimation,vansteelandt2014structural}. Under the restriction of effect stationarity (i.e., having common blip parameters across the measurement occasions), we can parameterize the blip as a simple function of the history  \cite[Section 5.1]{vansteelandt2014structural} that is
$\gamma_j^*(a_{j},\vec h_{ij};\vec\psi)=a_{j}\vec h_{ij}^\top\vec\psi$, where $\vec h_{ij}$ contains a one and potential confounders chosen from the histories. In this formulation, each component of $\vec\psi$  represents the change in the effect of treatment associated with the corresponding covariate. Under this semiparametric approach, the blip model needs to be correctly specified as a function of the history for consistent estimation of $\vec\psi$. We construct the proximal blipped down outcome as $U_{ij}=Y_{ij} - \gamma_j^*({A}_{ij}, \vec H_{ij};\vec\psi)$ at the $j$-th occasion.  These blipped down outcomes are a transformation of the observed data such that
\begin{align*}
    \text{E}(U_{ij}|\vec H_{ij}=\vec h_{ij},A_{ij}=a_{j})= \text{E}\{Y_{ij}(\overline{a}_{j-1}, 0)|\vec H_{ij}=\vec h_{ij},A_{ij}=a_{j}\}.
\end{align*}
This implies that 
the variable $U_{ij}$ has the same mean as the potential outcome under the reference treatment level  at occasion $j$. 
Under the above-mentioned assumptions, the efficient score function \citep{tsiatis2006semiparametric, chakraborty2013statistical} for $\vec\psi$ is given by
\begin{align} \label{eff.score}
    \vec S^{\text{eff}}(\vec\psi)&= \sum_{i=1}^n \Big[\frac{\partial\vec\gamma^*(\vec A_i,\vec H_i;\vec\psi)}{\partial\vec\psi^\top} - \text{E}\Big\{\frac{\partial\vec\gamma^*(\vec A_i,\vec H_i;\vec\psi)}{\partial\vec\psi^\top}|\vec H_i\Big\}\Big]^\top\,\text{Var}(\vec U_i|\vec H_i)^{-1}\{\vec U_i - \text{E}(\vec U_i|\vec H_i)\},
\end{align}
where $\vec A_i=(A_{i1},\ldots, A_{iJ})^\top$, $\vec H_i=(\vec H_{i1},\ldots,\vec H_{iJ})^\top$ is a $J\times K$ matrix representing the unit-wise history for the $i$-th subject, $\vec U_i = (U_{i1},\ldots,U_{iJ})^\top$, $\text{E}(\vec U_i|\vec H_i)= \vec H_i\vec\delta$ is the treatment-free model, and $\vec\delta$ denotes the corresponding parameters. 

We define the whole parameter vector as $\vec\theta=(\vec\delta^\top,\vec\psi^\top)^\top$ and denote the corresponding efficient score function by $\vec S^{\text{eff}}(\vec\theta)$ (computational form is provided in Section~\ref{form}). We express the covariance matrix of $\vec U_i$ as
\begin{align*}
    \text{Var}(\vec U_i|\vec H_i) = \vec Q_i^{1/2}\,\vec R_i(\alpha)\,\vec Q_i^{1/2},
\end{align*}
where $\vec U_i=\vec Y_i- \vec A_i\cdot\vec H_i\vec\psi$, $\vec Q_i=\sigma^2\vec I_{(J)}$ and $\vec R_i(\alpha)$ is the $J\times J$ matrix representing the correlations among the blipped down outcomes of a patient and is defined with parameter $\alpha$. The form of $\text{Var}(\vec U_i|\vec H_i)$ resembles the working covariance structure used in generalized estimating equations \citep{liang1986longitudinal}. Since $\vec Q_i$ and $\vec R_i(\alpha)$ are unknown, we replace these quantities in $\vec S^{\text{eff}}(\vec\theta)$ by their estimates. We estimate the parameters $\sigma^2$ and $\alpha$ using residual based moment method under a working correlation structure. We replace $\vec R_i(\alpha)$ by the estimated correlation matrix $\widehat{\vec R}$.

We define $\vec D_{ij} = [\vec H_{ij}^\top, \{A_{ij} - \text{E}(A_{ij}|\vec H_{ij})\}\cdot\vec H_{ij}^\top]^\top$ and $\vec D_i = (\vec D_{i1}, \ldots,\vec D_{iJ})^\top$ of dimension $J\times2K$ for $i=1,\ldots,n$, $j=1,\ldots,J$. With $\text{Var}(\vec U_i|\vec H_i)$ replaced with its estimate, we can write the efficient score as $\vec S^{\text{eff}}(\vec\theta) = \sum_{i=1}^n \vec S_i^{\text{eff}}(\vec\theta)$, where the contribution in the efficient score from the $i$-th subject can be expressed as
\begin{align*}
  \vec S_i^{\text{eff}}(\vec\theta) =  \vec D_{i}^\top\,\text{Var}(\vec U_i|\vec H_i)^{-1} \{\vec U_i - \text{E}(\vec U_i|\vec H_i)\}
  = \vec D_{i}^\top\,\widehat{\vec Q}_i^{-1/2}\,\widehat{\vec R}^{-1}\,\widehat{\vec Q}_i^{-1/2}\{\vec U_i - \text{E}(\vec U_i|\vec H_i)\}.
\end{align*}
The G-estimates $\widehat{\vec\theta}_n$ are obtained by solving the estimating equations
\begin{align} \label{unpenalized.score.equation}
   \vec S^{\text{eff}}(\vec\theta)=\vec 0. 
\end{align}

\subsection{Proposed method (Penalized G-estimation)} \label{sec.penalizedG}

For simultaneous selection of effect modifiers and estimation of the parameters, we propose a penalized efficient score function using a nonconvex smoothly clipped absolute deviation (SCAD) penalty \citep{fan2001variable}. The SCAD penalty achieves three desirable properties of variable selection: unbiasedness, sparsity, and continuity \citep{fan2001variable}. 
Under the assumptions for identifiability of the target parameter $\vec\psi$ given in section~\ref{notation}, we propose the following penalized efficient score function
\begin{align} \label{pen.score}
    \vec S^{P}(\vec\psi)&= \sum_{i=1}^n \Big[\frac{\partial\vec\gamma^*(\vec A_i,\vec H_i;\vec\psi)}{\partial\vec\psi^\top} - \text{E}\Big\{\frac{\partial\vec\gamma^*(\vec A_i,\vec H_i;\vec\psi)}{\partial\vec\psi^\top}|\vec H_i\Big\}\Big]^\top\,\text{Var}(\vec U_i|\vec H_i)^{-1} \nonumber \\
    &{\hspace{2.3in}}\times\{\vec U_i - \text{E}(\vec U_i|\vec H_i)\}- n{\vec q}_{\lambda_n}(|\vec\psi|)\text{sign}(\vec\psi),
\end{align}
where $\vec q_{\lambda_n}(|\vec\psi|)=\{0, q_{\lambda_n}(|\psi_1|),\ldots,q_{\lambda_n}(|\psi_{K-1}|)\}^\top$ and $q(.)$ indicates the first-derivative of the SCAD penalty function given by
\begin{align*}
    q_{\lambda_n}(\psi) = \lambda_n\Big\{I(\psi \leq\lambda_n)+\frac{(b\lambda_n-\psi)_+}{(b-1)\lambda_n}I(\psi >\lambda_n)\Big\}
\end{align*}
for $\psi \geq 0$ and some $b>2$ with $x_+=xI(x>0)$. \cite{fan2001variable} suggested to use $b=3.7$. The amount of shrinkage in estimation is determined by the tuning parameter $\lambda_n$.

For the whole parameter vector $\vec\theta$, the proposed penalized efficient score function becomes
\begin{align*}
  \vec S^{P}(\vec\theta) = \vec S^{\text{eff}}(\vec\theta) - n{\vec q}_{\lambda_n}(|\vec\theta|)\text{sign}(\vec\theta),
\end{align*}
where ${\vec q}_{\lambda_n}(|\vec\theta|) = \{\vec 0^\top,{\vec q}_{\lambda_n}(|\vec\psi^\top|)\}^\top$. The penalized estimates are obtained by solving the following estimating equations:
\begin{align} \label{penalized.score.equation}
   \vec S^{P}(\vec\theta)=\vec 0. 
\end{align}
Note that we do not penalize the parameters $\vec\delta$ of the treatment-free model and the main effect of the treatment ($\psi_0$), because our goal is only to identify the effect modifiers. 

For solving the penalized efficient estimating equations (\ref{penalized.score.equation}), we use the minorization-maximization (MM) algorithm for nonconvex penalty discussed by \cite{hunter2005variable}. We combine the MM algorithm with the iterative procedure of G-estimation. According to the MM algorithm, the penalized G-estimator $\widetilde{\vec\theta}$ approximately satisfies
\begin{align} \label{MM.score}
   S_{m}^{\text{eff}}(\widetilde{\vec\theta}) - n\, q_{\lambda_n}(|\widetilde{\theta}_m|)\text{sign}(\widetilde{\theta}_m)\frac{|\widetilde{\theta}_m|}{\epsilon+|\widetilde{\theta}_m|} = 0 
\end{align}
for $m=1,\ldots 2K$, where $S_{m}^{\text{eff}}(\widetilde{\vec\theta})$ denotes the $m$-th element of $\vec S^{\text{eff}}(\widetilde{\vec\theta})$ and $\epsilon$ can be a small number, e.g.\ $10^{-6}$. If we take a closer look, it is easy to understand that equations (\ref{MM.score}) are perturbed versions of the original estimating equations (\ref{penalized.score.equation}). For obtaining the penalized G-estimates, the Newton-Raphson type updating formula at the $t$-th iteration directly follows from equation (\ref{MM.score}) and is given by:
\begin{align} \label{update.penalizedG}
    \vec\theta^{t+1} = \vec\theta^{t} + \Big\{\vec H_n(\vec\theta^{t})+n\vec E_n(\vec\theta^{t})\Big\}^{-1}\Big\{\vec S^{\text{eff}}(\vec\theta^{t})-n\vec E_n(\vec\theta^{t})\vec\theta^{t}\Big\},
\end{align}
where 
\begin{align} \label{eq.Hn}
     \vec{H}_n(\vec\theta^{t}) = -\text{E}\Big(\dfrac{\partial\vec{S}^\text{eff}(\vec\theta)}{\partial\vec\theta^\top}\Big)\Big|_{\vec\theta=\vec\theta^t}
\end{align}
and
\begin{align} \label{eq.En}
\vec{E}_n(\vec\theta^{t}) = \text{diag}\Bigg\{\vec{0}^\top,\frac{q_{\lambda_n}(|\psi_1|)}{\epsilon+|\psi_1|},\ldots,\frac{q_{\lambda_n}(|\psi_{K-1}|)}{\epsilon+|\psi_{K-1}|}\Bigg\} \Bigg|_{\vec\psi=\vec\psi^t}.
\end{align}

In equation (\ref{update.penalizedG}), the derivative of the efficient score, $\vec{H}_n(\vec\theta^{t})$, is a $2K \times 2K$-dimensional matrix and $\vec{E}_n(\vec\theta^{t})$ is a $2K \times 1$-dimensional vector. The performance of the penalized estimators rely on the proper choice of tuning parameter. We use a doubly robust information criterion (DRIC) \citep{bian2023variable,moodie2023variable} for selecting the tuning parameter (technical details regarding this criterion are provided in Section~\ref{form} of the Appendix). For a specific correlation structure (corstr), the steps for the whole estimation procedure are summarized in Algorithm~\ref{penG.algorithm}.

\begin{algorithm}
\caption{Penalized G-estimation algorithm}\label{penG.algorithm}
\begin{algorithmic}[1]
\Procedure{penalizedG}{$\vec A, \vec H, \vec Y, \lambda$,\,corstr,\,$\kappa$} 
    \State Compute $\text{E}(\vec A_i|\vec H_i)$ for $i=1,\ldots,n$ 
    \State $\vec\theta^{\text{up}} \gets \{\sum_{i=1}^n\vec D_i^\top(\vec H_i\;\;\vec A_i\cdot\vec H_i)\}^{-1}\sum_{i=1}^n\vec D_i^\top \vec Y_i$ 
    \ForAll{$\lambda \in (\lambda_{\text{max}}, \ldots, \lambda_{\text{min}})$} 
        \State Initialize: $t = 0$, $\vec\theta^0 \gets \vec\theta^{\text{up}}$
            \Repeat
                \State $\vec e_i \gets \vec Y_i - (\vec H_i\;\;\vec A_i\cdot\vec H_i)\vec\theta^t$ for $i=1,\ldots,n$
                \State Compute $\sigma^t$ and $\alpha^t$ under corstr
                \State Compute $\widehat{\vec V}_i$  according to corstr for $i=1,\ldots,n$
                \State Compute $\vec S^{\text{eff}}(\vec\theta^t)$
                \State Compute $\vec H_n(\vec\theta^t)$ and $\vec E_n(\vec\theta^t)$ using (\ref{eq.Hn}) and (\ref{eq.En}).
                \State Update $\vec\theta^{t}$ according to (\ref{update.penalizedG}) and obtain $\vec\theta^{t+1}$
                \State $t \gets t+1$
            \Until{$||\vec\theta^{t} - \vec\theta^{t-1}|| < \kappa$}
        \State $\widetilde{\vec\theta}_\lambda \gets \vec\theta^t$, $\widetilde{\sigma}_\lambda \gets \sigma^t$, and $\widetilde{\alpha}_\lambda \gets \alpha^t$
        \State Compute $\text{DRIC}_{\lambda}$
    \EndFor
    \State $\widetilde{\vec\theta} \gets \widetilde{\vec\theta}_\lambda^*$ s.t.\ $\widetilde{\vec\theta}_\lambda^*$ corresponds to the minimum of $\text{DRIC}_{\lambda}$
    \State \textbf{return} $\widetilde{\vec\theta}$
\EndProcedure
\end{algorithmic}
\end{algorithm}
In Algorithm~\ref{penG.algorithm}, $\text{E}(\vec A_i|\vec H_i)$ in Step 2 is based on a logistic regression on the pooled data and the initial estimator $\vec\theta^{\text{up}}$ in Step 3 is the univariate unpenalized G-estimator. The computational formulas for the method of moments estimators $\sigma^t$ and $\alpha^t$ of Step 8, $\vec S^{\text{eff}}(\vec\theta^t)$ of Step 10 and $\text{DRIC}_{\lambda}$ of Step 16 are provided in Sections~\ref{moment.method},~\ref{form} and~\ref{tune.par.selection} of the Appendix, respectively. We set $\kappa=10^{-6}$ in Step 14. Details regarding computation of the correlation matrix $\widehat{\vec R}_i(\alpha)$ in $\widehat{\vec V}_i$ under different structures can be found in \cite{sultana2023caution}. Note that, $\widehat{\vec V}_i=\widehat{\vec V}$  $\forall i$ if $J$ is fixed.

\subsection{Asymptotic properties} \label{sec.asymp}

In this section, we study the asymptotic properties of the proposed penalized G-estimator where we assume a constant cluster size (i.e. the same numbers of sessions for patients).  Let $\vec\theta_0=(\vec\delta_0^\top,\vec\psi^\top_0)^\top$ have dimension $2K$, where $\vec\delta_0$ is the $K$-dimensional vector of population parameters corresponding to the assumed treatment-free model and $\vec\psi_0$ is the $K$-dimensional vector of true coefficients of the blip model. Let $B_1 = \{1,\ldots,K+1\}$ be the set representing the indices of the unpenalized main effects ($\vec \delta_0$ and $\psi_{00}$) and $B_2 = \{m: \psi_{0m}\neq 0; m=K+2, \ldots,2K\}$ be the index set for the variables that are true  effect modifiers of the treatment within this model. Let $B=B_1 \cup B_2$ and $s$ denote the cardinality of the set $B$. Then let $B^c$ represent the indices of the zero coefficients in $\vec\psi_0$, i.e.\ the indices of the variables that do not modify the effect of the treatment.

Let $\widehat{\vec R}$ introduced in Section~\ref{sec.G.estimation} satisfy 
$||\widehat{\vec R}^{-1} - \overline{\vec R}^{-1}|| = O_p(\sqrt{1/n})$, 
where $\overline{\vec R}=\text{E}(\widehat{\vec R})$ is a constant positive definite matrix with eigenvalues bounded away from zero and $+\infty$. Note that $||\vec X||=\{trace(\vec X\vec X^\top)\}^{1/2}$ denotes the Frobenius norm of the matrix $\vec X$. 
Define two versions of the expected information as 
\begin{align*}
\vec H(\vec\theta)&= \text{E}\{-\partial\vec S_i^{\text{eff}}(\vec\theta)/\partial\vec\theta^\top\} = \vec D_{i}^\top\vec Q_i^{-1/2}\overline{\vec R}^{-1}\vec Q_i^{-1/2}(\vec H_i\;\;\vec A_i\cdot\vec H_i),\\
\vec I(\vec\theta)&= \text{E}\{\vec S_i^{\text{eff}}(\vec\theta)\vec S_i^{\text{eff}}(\vec\theta)^\top\} = \vec D_{i}^\top\vec Q_i^{-1/2}\overline{\vec R}^{-1}\vec R_0\overline{\vec R}^{-1}\vec Q_i^{-1/2}\vec D_{i},
\end{align*}
where $\vec R_0$ is the true correlation matrix. Details regarding the expressions of these information matrices can be found in \cite{balan2005asymptotic}, where the authors presented a rigorous asymptotic theory for generalized estimating equations.

According to \cite{johnson2008penalized}, for establishing the asymptotic properties, the required conditions on the efficient score function and the penalty function are as follows:
\begin{enumerate}
    \item [(C.a)] There exists a non-singular matrix $\vec H(\vec\theta_0)$ such that for any given constant $Z$,
    \begin{align*}
        \sup_{|\theta-\theta_0| \leq Zn^{-1/2}} |n^{-1/2}\vec S^{\text{eff}}(\theta)-n^{-1/2}\vec S^{\text{eff}}(\theta_0)-n^{1/2}\vec H(\vec\theta_0)(\theta-\theta_0)| = o_p(1).
    \end{align*}
    Moreover,
    \begin{align*}
     n^{-1/2}\vec S^{\text{eff}}(\vec\theta_0) \xrightarrow[]{d} N(\vec 0, \vec I(\vec\theta_0)).
    \end{align*}
    \item [(C.b)]The derivative $q_{\lambda_n}(\cdot)$ of the penalty function has the following properties:
    \begin{enumerate}
        \item [(C.b.1)] For nonzero fixed $\theta$, $\lim n^{1/2}q_{\lambda_n}(|\theta|)=0$ and $\lim q'_{\lambda_n}(|\theta|)=0$.
        \item [(C.b.2)] For any $Z > 0$, $\lim \sqrt{n}\inf_{|\theta| \leq Zn^{-1/2}}q_{\lambda_n}(|\theta|) \rightarrow \infty$,
    \end{enumerate}
\end{enumerate}
where $q'_{\lambda_n}(\cdot)$ denotes the derivative of $q_{\lambda_n}(\cdot)$. 
Under the conditions C.a, C.b and the regularity conditions C1-C7 (provided in Section~\ref{regularity.cond} of the Appendix) applicable for GEE, there exists a $\sqrt{n}$-consistent approximate zero crossing of $\vec  S^P(\vec\theta)$ indicating $\widetilde{\vec\theta}_n=\vec\theta_0+O_p(n^{-1/2})$, such that $\widetilde{\vec\theta}_n$ is an approximate zero crossing of $\vec  S^P(\vec\theta)$. The proof is analogous to what is presented in \cite{johnson2008penalized}. Also, the proposed estimator enjoys the oracle property, which means that it estimates true zero coefficients as zero with probability approaching one and the true nonzero coefficients as efficiently as if the true model is known in advance. This property is stated in the next two theorems.

\begin{theorem}[Correct Sparsity] \label{thm1}
Let $\vec\theta^B$ represent the elements of $\vec\theta$ whose indices belong to $B$. Under the required conditions (C.a, C.b and C1-C7), if $\lambda_n \rightarrow 0$ and $\sqrt{n}\lambda_n \rightarrow \infty$ as $n \rightarrow \infty$, then there exists an approximate penalized G-estimator $\widetilde{\vec\theta}_n=\{\widetilde{\vec\theta}_n^{B},\widetilde{\vec\theta}_n^{B^c}\}$ such that
\begin{align} \label{theorem1}
    P\Big(\widetilde{\vec\theta}_n^{B^c}=0\Big) \rightarrow 1.
\end{align}   
\end{theorem}
This theorem establishes the correct sparsity of the penalized G-estimator. The proof of Theorem~\ref{thm1} is provided in Section~\ref{proof.thm1} of the Appendix.

\begin{theorem}[Asymptotic Normality] \label{thm2}
  Let $\vec D_{i,B}$ represent the elements of the matrix $\vec D_i$ that correspond to the coefficients $\vec\theta^B$. Under the required conditions (C.a, C.b and C1-C7), if $\lambda_n \rightarrow 0$ and $\sqrt{n}\lambda_n \rightarrow \infty$ as $n \rightarrow \infty$, the penalized G-estimator $\widetilde{\vec\theta}_n^{B}$ satisfies: 
\begin{align*}
    &n^{1/2}\Big\{\vec H_{B}(\vec\theta_0)+\vec W_n(\vec\theta^B_0)\Big\}\Big[\widetilde{\vec\theta}_n^B-\vec\theta^B_0+\Big\{\vec H_{B}(\vec\theta_0)+\vec W_n(\vec\theta^B_0)\Big\}^{-1}\vec b_n\Big]
    \xrightarrow[]{d} N\Big(\vec 0, \vec I_B(\vec\theta_0)\Big).
\end{align*}
where $\vec H_{B}(\vec\theta_0)$, $\vec W_{n}(\vec\theta_0^B)$ and $\vec I_{B}(\vec\theta_0)$ are the $s\times s$ submatrices of $\vec H(\vec\theta_0)$, $\vec W_{n}(\vec\theta_0)$ and $\vec I(\vec\theta_0)$ that correspond to the indices in $B$,
\begin{align*}
 \vec W_{n}(\vec\theta_0) = \text{diag}\{\vec 0^\top,-q_{\lambda_n}'(|\psi_{01}|)\text{sign}(\psi_{01}),\ldots,-q_{\lambda_n}'(|\psi_{0(K-1)}|)\text{sign}(\psi_{0(K-1)})\},
\end{align*}
and $\vec b_n = -\vec q_{\lambda_n}(|\vec\theta_0^B|)\text{sign}(\vec\theta_0^B)$.  
\end{theorem}
This theorem establishes the asymptotic normality of the penalized G-estimator $\widetilde{\vec\theta}_{n}^B$. The proof of Theorem~\ref{thm2} is provided in Section~\ref{proof.thm2} of the Appendix.
\begin{corollary}[Double Robustness]
    If the consistency and the sequential ignorability assumptions in Section~\ref{notation} are correct, the penalized G-estimator $\widetilde{\vec\psi}_n$ is doubly-robust, i.e., $\widetilde{\vec\psi}_n$ is a consistent estimator for $\vec\psi$ if either the exposure model or the treatment-free model is correctly specified given that the candidate set for the blip function contains the true blip model. It follows from Theorems~\ref{thm1} and~\ref{thm2} that penalized G-estimator inherits the double-robustness property of G-estimator.
\end{corollary}

Although the asymptotic properties are derived under the same number of measurements per subject for the ease of proof, our theory and related estimator can handle unequal numbers of measurements. In the simulation study of Section~\ref{sec.results1}, we numerically verified that the results of the theorems also hold for non-fixed sessions.

An asymptotic sandwich covariance estimator for $\widetilde{\vec\theta}_n^B$ directly follows from Theorem~\ref{thm2} and is given by 
\begin{align} \label{sand.penalized.thetaB}
    \text{Cov}(\widetilde{\vec\theta}_n^B) = n^{-1}\Big\{\vec H_{B}(\vec\theta_0)+\vec W_n(\vec\theta^B_0)\Big\}^{-1} \vec I_B(\vec\theta_0) \Big\{\vec H_{B}(\vec\theta_0)+\vec W_n(\vec\theta^B_0)\Big\}^{-1}.
\end{align}

We can consistently estimate the variance in (\ref{sand.penalized.thetaB}) by replacing the quantities in it by their empirical estimates as follows
\begin{align*}
\widehat{\text{Cov}}(\widetilde{\vec\theta}_n^B) = n^{-1}\Big\{\widehat{\vec H}_{B}(\widetilde{\vec\theta}_n)+\vec W_n(\widetilde{\vec\theta}_n^B)\Big\}^{-1} \widehat{\vec I}_B(\widetilde{\vec\theta}_n) \Big\{\widehat{\vec H}_{B}(\widetilde{\vec\theta}_n)+\vec W_n(\widetilde{\vec\theta}_n^B)\Big\}^{-1} 
\end{align*}
where
 $   \widehat{\vec I}_B(\widetilde{\vec\theta}_n) = n^{-1}\sum_{i=1}^n \vec S_i^{\text{eff}}(\widetilde{\vec\theta}_n^B)\Big\{\vec S_i^{\text{eff}}(\widetilde{\vec\theta}_n^B)\Big\}^\top$
and
$    \widehat{\vec H}_{B}(\widetilde{\vec\theta}_n)= -n^{-1}\sum_{i=1}^n \dfrac{\partial S_i^{\text{eff}}(\vec\theta)}{\partial\vec\theta^\top}\Big |_{\vec\theta=\widetilde{\vec\theta}_n^B}$.
\begin{remark}
Theorem~\ref{thm2} requires the propensity score (i.e., treatment model parameters) to be known. Because the propensity score is unknown and must be estimated using the observed data, the variance of our estimator should be calculated as shown in Section~\ref{appendix.sandwich} of the Appendix. This section shows how we can modify our estimating equation to account for the contribution of the estimated propensity score in the asymptotic variance of our estimator. This approach was adapted from \cite{robins1992estimation}.
\end{remark}
\begin{remark}
The sandwich variance estimator is valid if we have infinitely large samples. However, in finite samples, the uncertainty in the selection of effect modifiers invalidates the post-selection inference based on this sandwich estimator. Under data-driven selection, we tend to favour models with strong effects with an associated cost of inflated type I errors \citep{zhao2022selective}. Appropriate statistical approach is required for a valid post-selection inference for the penalized G-estimation in finite samples.
\end{remark}
\section{Simulation study} \label{simulation}

In order to evaluate the performance of the proposed estimator in finite samples we conducted two simulation studies. The first evaluates the oracle property from Theorem~\ref{thm1} and Theorem~\ref{thm2}; the setup and results are presented in Section~\ref{sec.results1}. The second simulation study in Section~\ref{sec.double.robust} verifies the double robustness property of our estimator.

\subsection{Evaluation of the oracle property in finite samples} \label{sec.results1}
To generate the data for $j$-th session ($j=1,\ldots,J$) of each subject, we generated two baseline confounders as $L^{(1)} \sim Bernoulli(0.5)$ and $L^{(2)} \sim N(0,1)$, and the time varying confounders and noise covariates as $L_j^{(3)},\ldots,L_j^{(6)},X_j^{(1)},\ldots,X_j^{(10)} \sim MVN_{14}\Big((\vec\mu_{L,j}^\top,\vec\mu_{X,j}^\top)^\top,\vec V\Big)$, where $\mu_{L,j}^{(k)}= 0.3\,l^{(k)}_{j-1}+0.3\,a_{j-1}$ for $k=3,4,5 \text{ and } 6$, and $\mu_{X,j}^{(r)}= 0.5\,x^{(r)}_{j-1}$ for $ r=1,\ldots,10$. Note that we define $l_0^{(k)} =0$ for any $k$, $a_0 =0$, and $x_0^{(r)} =0$ for any $r$. The covariance matrix $\vec V$ has $(r,s)$-th element equal to $\rho^{|r-s|}$ for $r,s=1,\ldots,14$. We generated the binary exposure according to the probability
\begin{align}
    P(A_j = 1 | \vec H_j) = \dfrac{\exp{(\beta_0+\beta_1 l^{(1)}+\beta_2 l^{(2)}+\sum_{m=3}^6 \beta_m l_j^{(m)} + \beta_7 a_{j-1})}}{1+\exp{(\beta_0+\beta_1 l^{(1)}+\beta_2 l^{(2)}+\sum_{m=3}^6 \beta_m l_j^{(m)} + \beta_7 a_{j-1})}}.
\end{align}
We then generated a vector of correlated errors $\vec \epsilon \sim N_{J}(\vec 0, \vec \Sigma)$, where $\vec \Sigma=\sigma^2_\epsilon\vec R$ is the variance-covariance matrix and $\vec R$ is the $J\times J$ correlation matrix defined with parameter $\alpha$ according to a ``exchangeable" correlation structure. Finally, we constructed the outcome as 
\begin{align*}
    y_j = \mu_j(\vec h_j; \vec\delta) + \gamma^*_j(a_j,\vec h_j; \vec\psi) + \epsilon_j,
\end{align*}
where
$\mu_j(\vec h_j; \vec\delta)=\delta_0+\delta_1 l^{(1)}+\delta_2 l^{(2)}+\sum_{m=3}^6 \delta_m l_j^{(m)}+\delta_7 \exp(l_j^{(5)})+ \delta_8 a_{j-1}$ and $\gamma^*_j(a_j,\vec h_j; \vec\psi)=(\psi_0+\psi_1 l^{(1)}+\psi_2 l^{(2)}+\sum_{m=3}^6\psi_m l_j^{(m)}+ \psi_7 a_{j-1})a_j$.
Let $\vec\beta=(\beta_0,\ldots,\beta_7)^\top$, $\vec\delta=(\delta_0,\ldots,\delta_8)^\top$ and $\vec\psi=(\psi_0,\ldots,\psi_7)^\top$. We fixed $\vec\beta=(0,1,1,1,1,1,1,-0.8)^\top$, $\vec\delta=(1,-1,1,1,1,1,1,1,1)^\top$ and we considered two different models for the blip;
\begin{itemize}
    \item {\bf Setting 1} (stronger effect modification): $\vec\psi=(1, -2.5, 1.5, 1.5, 1.5, 1.5, 0, 2)^\top$
    \item {\bf Setting 2} (weaker effect modification): $\vec\psi=(1, -2, 1, 0.75, 0.9, 1.2, 0, 1.8)^\top$.
\end{itemize}
Note that none of the $X$'s were used to generate $\mu_j(\vec h_j; \vec\delta)$ and $\gamma^*_j(a_j,\vec h_j; \vec\psi)$. 
We considered $n=200$, $\rho=0$ vs.\ 0.25, $\sigma^2_\epsilon=1$ vs.\ 4 and $\alpha=0.8$. All of the confounders and noise covariates were considered candidates for effect modification. We performed the proposed penalized G-estimation using three different working correlation structures: independent (Indep), exchangeable (Exch), and unstructured (UN). In all the scenarios, we compare the estimates under a misspecified treatment-free model. We consider a variable to be eliminated by the method if $|\text{effect modification}| < 0.001$
. For evaluating the model selection performance, we considered the rate of false negatives (FN) (i.e.\ the proportion of times the method eliminated at least one true effect modifier), the rate of false positives (FP) (i.e.\ the proportion of times the method select at least one covariate which is not a true effect modifier), the rate of exact selections (EXACT) (i.e.\ proportion of times the method selected the true set of effect modifiers), and the average number of false positives (AFP) (i.e.\ the average number of variables selected by the method which do not belong to the set of true effect modifiers). The performance measures for different settings were calculated from 500 independent simulations and are shown in Table~\ref{tab1}.

\begin{table}[ht]
\centering
\caption{Model selection performance of the penalized G-estimator for data generated with $n=200$, $J=6$ for all $i$, an exchangeable correlation structure and $\alpha=0.8$ under varying values of the error variance $\sigma^2_\epsilon$ and varying degrees of autocorrelation between the effect modifiers and noise covariates. Results are obtained with a misspecified treatment-free model from 500 independent simulations for Setting 1 and Setting 2.}
\begin{tabular}{ccccccccccc}
  \toprule
  Autocorr& Working & \multicolumn{4}{c}{\bf Setting 1 (Stronger EM)}&& \multicolumn{4}{c}{\bf Setting 2 (Weaker EM)}\\
  \cline{3-6}\cline{8-11}
coeff ($\rho$)& correlation & FN & FP & EXACT & AFP &  & FN & FP & EXACT & AFP \\ 
  \hline\hline
   && \multicolumn{4}{c}{$\sigma^2_\epsilon=1$}&& \multicolumn{4}{c}{$\sigma^2_\epsilon=1$}\\
   \cline{3-6}\cline{8-11}
& Indep & 2.0 & 1.4 & 96.6 & 0.01 &  & 9.2 & 3.6 & 87.8 & 0.04 \\ 
 0  & \bf Exch & 1.6 & 1.4 & \bf 97.0 & 0.01 &  & 8.8 & 2.8 & \bf 88.8 & 0.03 \\ 
   & UN & 1.4 & 1.8 & 96.8 & 0.02 &  & 10.0 & 2.6 & 87.4 & 0.03\vspace{.2cm}\\ 
   & Indep & 2.2 & 6.0 & 91.8 & 0.07 &  & 3.6 & 6.6 & 90.2 & 0.07 \\ 
 0.25  & \bf Exch & 1.6 & 5.6 & \bf 92.8 & 0.06 &  & 2.4 & 5.8 & \bf 92.2 & 0.06 \\ 
   & UN & 1.0 & 4.8 & 94.2 & 0.06 &  & 2.0 & 3.8 & 94.4 & 0.04 \\    
   \hline
  && \multicolumn{4}{c}{$\sigma^2_\epsilon=4$}&& \multicolumn{4}{c}{$\sigma^2_\epsilon=4$}\\
  \cline{3-6}\cline{8-11}
  & Indep & 14.8 & 2.4 & 83.0 & 0.02 &  & 44.6 & 1.4 & 54.0 & 0.02 \\ 
 0  & \bf Exch & 14.2 & 1.0 & \bf 84.8 & 0.01 &  & 42.4 & 0.4 & \bf 57.2 & 0.00 \\ 
   & UN & 14.6 & 0.0 & 85.4 & 0.00 &  & 45.8 & 0.2 & 54.2 & 0.00\vspace{.2cm}\\ 
   & Indep & 11.8 & 3.8 & 84.4 & 0.04 &  & 28.2 & 3.4 & 69.0 & 0.03 \\ 
  0.25 & \bf Exch & 10.6 & 0.6 & \bf 88.8 & 0.01 &  & 25.4 & 1.6 & \bf 73.2 & 0.02 \\ 
   & UN & 9.4 & 0.6 & 90.0 & 0.01 &  & 24.8 & 1.6 & 74.0 & 0.02 \\  
   \bottomrule
\end{tabular} \label{tab1}
\vspace{0.1cm}\\
  \footnotesize
FN: \% of false negatives, FP: \% of false positives, EXACT: \% of exact selections, AFP: average false positives,\\ EM: effect modifier, Indep: independent, Exch: exchangeable, UN: unstructured  
\end{table}

The selection rates of the exact model are above 90\% for most cases when the error variance is one. When the error variance was increased, the exact selection rate decreased and the false negative rate increased. When the outcomes were more highly correlated and the candidate variables were auto-correlated among themselves, accounting for correlation among the observed outcomes within a patient produced a clear advantage over the independence assumption. In this case, the false negative and false positive rates produced by the working independent structure are higher than the other two structures. 

Simulations were also conducted for a larger sample size by setting $n=500$ and two different values for the correlation parameter by setting $\alpha=0.4$ and 0.8 indicating low and high correlation among the repeated outcomes, respectively; the results are presented in Table~\ref{large.n.tab}. Under the larger sample size ($n=500$) with all other parameters being the same, we see an improved performance of the proposed method (compare the second row of Table~\ref{tab1} to the bottom row of Table~\ref{large.n.tab}), supporting the result of Theorem~\ref{thm1}.

\begin{table}[ht]
\centering
\caption{Model selection performance of the penalized G-estimator for data generated with $n=500$, $J=6$ for all $i$, an exchangeable correlation structure among the repeated outcomes with varying value of correlation parameter $\alpha$, an autocorrelation coefficient $\rho=0.25$ deciding the correlation among the EM's and noise covariates, and error variance $\sigma^2_\epsilon=1$. Results are obtained with a misspecified treatment-free model from 500 independent simulations for Setting 1 and Setting 2.}
\begin{tabular}{ccccccccccc}
  \toprule
& Working & \multicolumn{4}{c}{\bf Setting 1 (Stronger EM)}&& \multicolumn{4}{c}{\bf Setting 2 (Weaker EM)}\\
  \cline{3-6}\cline{8-11}
$\alpha$& correlation & FN & FP & EXACT & AFP &  & FN & FP & EXACT & AFP \\ 
  \hline\hline
  & Indep & 0.4 & 7.2 & 92.4 & 0.08 &  & 0.0 & 5.4 & 94.6 & 0.05 \\ 
 0.4  & \bf Exch & 0.4 & 7.2 & \bf 92.4 & 0.08 &  & 0.0 & 5.8 & \bf 94.2 & 0.06 \\ 
   & UN & 0.4 & 5.6 & 94.0 & 0.06 &  & 0.0 & 4.6 & 95.4 & 0.05\vspace{.2cm}\\ 
   & Indep & 0.0 & 5.8 & 94.2 & 0.06 &  & 0.2 & 8.8 & 91.0 & 0.09 \\ 
 0.8  & \bf Exch & 0.0 & 5.2 & \bf 94.8 & 0.06 &  & 0.2 & 6.6 & \bf 93.2 & 0.07 \\ 
   & UN & 0.0 & 4.6 & 95.4 & 0.05 &  & 0.0 & 6.4 & 93.6 & 0.07 \\ 
   \bottomrule
\end{tabular} \label{large.n.tab}
\vspace{0.1cm}\\
\footnotesize
FN: \% of false negatives, FP: \% of false positives, EXACT: \% of exact selections, AFP: average false positives,\\ EM: effect modifier, Indep: independent, Exch: exchangeable, UN: unstructured 
\end{table}

We also compared the performance of the penalized estimator with the Oracle estimator (estimator obtained from the true blip model but with the same misspecification in the treatment-free part) under each of the working correlation structures. In Table~\ref{sidewaytab}, we reported the relative bias, the empirical root-mean-squared error (SE1) and the square root of the average of sandwich variance estimates (SE2) under different working correlation structures. From the results, we see that the usage of non-independent correlation structures  resulted in more efficient estimation of the blip parameters. Moreover, the efficiency of the penalized estimates is comparable to that of the oracle estimates, which also verifies the property stated in Theorem~\ref{thm2}.

\begin{sidewaystable}
\centering
\caption{Relative bias (\%), empirical (SE1) and sandwich (SE2) estimates of the standard errors for the penalized  and the oracle estimators calculated from 500 simulations for $n=500$, autocorrelation coefficient $\rho = 0.25$, error variance $\sigma_\epsilon^2 = 1$, and $\alpha=0.8$ under the two settings.}
\begin{tabular}{ccccccccccccccccc}
  \toprule
  Working &  & \multicolumn{7}{c}{\bf Penalized} && \multicolumn{7}{c}{\bf Oracle}  \\ 
  \cline{3-9}\cline{11-17}
  model& Stats & $\psi_0$ & $\psi_1$ & $\psi_2$ & $\psi_3$ & $\psi_4$ & $\psi_5$ &$\psi_7$ && $\psi_0$ & $\psi_1$ & $\psi_2$ & $\psi_3$ & $\psi_4$ & $\psi_5$&$\psi_7$ \\ 
  \hline
  \bf Setting 1\\
  \bf (Stronger EM)\\
 Ind & r.bias & 1.32 & -1.04 & 0.05 & 0.00 & 0.40 & 0.06 & 0.21 &  & 1.37 & -0.94 & 0.13 & 0.18 & 0.23 & 0.11 & 0.04 \\ 
  Ind & SE1 & 0.16 & 0.19 & 0.11 & 0.11 & 0.11 & 0.18 & 0.19 &  & 0.16 & 0.19 & 0.11 & 0.11 & 0.11 & 0.18 & 0.19 \\ 
  Ind & SE2 & 0.16 & 0.19 & 0.11 & 0.10 & 0.11 & 0.18 & 0.20 &  & 0.16 & 0.19 & 0.11 & 0.10 & 0.11 & 0.18 & 0.20 \\ 
  Exch & r.bias & 1.13 & -1.11 & 0.01 & 0.06 & 0.44 & 1.30 & 0.24 &  & 1.17 & -1.03 & 0.16 & 0.21 & 0.30 & 1.35 & 0.03 \\ 
  Exch & SE1 & 0.14 & 0.18 & 0.10 & 0.10 & 0.10 & 0.17 & 0.17 &  & 0.14 & 0.18 & 0.10 & 0.10 & 0.10 & 0.17 & 0.17 \\ 
  Exch & SE2 & 0.14 & 0.18 & 0.10 & 0.10 & 0.10 & 0.17 & 0.18 &  & 0.14 & 0.18 & 0.10 & 0.10 & 0.10 & 0.17 & 0.18 \\ 
  UN & r.bias & 1.00 & -1.02 & 0.05 & 0.01 & 0.47 & 1.76 & 0.10 &  & 1.02 & -0.95 & 0.17 & 0.11 & 0.35 & 1.80 & 0.07 \\ 
  UN & SE1 & 0.15 & 0.19 & 0.11 & 0.10 & 0.10 & 0.17 & 0.18 &  & 0.15 & 0.19 & 0.11 & 0.10 & 0.10 & 0.17 & 0.18 \\ 
  UN & SE2 & 0.14 & 0.18 & 0.10 & 0.10 & 0.10 & 0.17 & 0.19 &  & 0.14 & 0.18 & 0.10 & 0.10 & 0.10 & 0.17 & 0.19 \\ 
    \bf Setting 2\\
    \bf (Weaker EM)\\
Ind & r.bias & 0.53 & -1.01 & 0.84 & 1.97 & 1.59 & 0.23 & 2.23 &  & 0.47 & -0.78 & 0.44 & 1.40 & 1.10 & 0.31 & 1.80 \\ 
  Ind & SE1 & 0.15 & 0.20 & 0.11 & 0.11 & 0.11 & 0.19 & 0.21 &  & 0.15 & 0.20 & 0.11 & 0.11 & 0.11 & 0.19 & 0.21 \\ 
  Ind & SE2 & 0.16 & 0.19 & 0.11 & 0.11 & 0.11 & 0.18 & 0.20 &  & 0.16 & 0.19 & 0.11 & 0.11 & 0.11 & 0.18 & 0.20 \\ 
  Exch & r.bias & 0.19 & -1.09 & 0.69 & 1.69 & 1.17 & 1.58 & 1.76 &  & 0.15 & -0.94 & 0.39 & 1.30 & 0.81 & 1.62 & 1.44 \\ 
  Exch & SE1 & 0.14 & 0.19 & 0.11 & 0.10 & 0.11 & 0.18 & 0.19 &  & 0.14 & 0.19 & 0.11 & 0.10 & 0.10 & 0.18 & 0.19 \\ 
  Exch & SE2 & 0.14 & 0.18 & 0.10 & 0.10 & 0.10 & 0.17 & 0.18 &  & 0.14 & 0.18 & 0.10 & 0.10 & 0.10 & 0.17 & 0.18 \\ 
  UN & r.bias & 0.61 & -0.93 & 0.85 & 1.46 & 1.13 & 2.28 & 1.82 &  & 0.57 & -0.76 & 0.53 & 1.04 & 0.75 & 2.38 & 1.49 \\ 
  UN & SE1 & 0.14 & 0.20 & 0.11 & 0.10 & 0.11 & 0.18 & 0.20 &  & 0.14 & 0.20 & 0.11 & 0.10 & 0.11 & 0.18 & 0.20 \\ 
  UN & SE2 & 0.14 & 0.18 & 0.10 & 0.10 & 0.10 & 0.17 & 0.19 &  & 0.14 & 0.18 & 0.10 & 0.10 & 0.10 & 0.17 & 0.19 \\ 
\bottomrule
\end{tabular} \label{sidewaytab}
\vspace{0.1cm}\\
\footnotesize
relative bias (r.bias) $=100\times (\widetilde{\psi}_n-\psi_0)/\psi_0$, SE1: the empirical root-mean squared error (MSE), SE2: the square-root of the average of sandwich variance estimates, Ind: independent, Exch: exchangeable, UN: unstructured 
\end{sidewaystable}

We performed additional simulations allowing for the number of visits (or measurement occasions) for each subject to be random; the results are provided in Table~\ref{unequal.J.tab} of the Appendix. We also considered alternative choices for the exposure generating model, for example, when the exposure selection probability depends on past outcome, and for the outcome generating model where past outcomes affect future outcomes. The results are provided in Tables~\ref{supp.tab.dag2} and~\ref{supp.tab.dag3}, respectively, in the Appendix. We performed additional simulations in a high-dimensional setup considering two different dimensions of covariates: $K=50$ vs.\ 100; the results are provided in Table~\ref{supp.tab.high} of the Appendix. We also compared our penalized G-estimator with an existing alternative, which is the proximal treatment effect estimator of \cite{boruvka2018assessing}, under the full model as Boruvka's method does not allow for variable selection; the results are presented in Table~\ref{supp.tab.boruvka} of the Appendix.

\subsection{Verification of the double robustness property} \label{sec.double.robust}

The double robustness of the proposed penalized G-estimator was verified via a simulation experiment with $n=500$ and $J=6$ for all subjects. We considered estimation under four different scenarios: {\bf Scenario 1}: The treatment model is correct, and the treatment-free model is misspecified, {\bf Scenario 2}: The treatment model is incorrect, and the treatment-free model is correct, {\bf Scenario 3}: Both models are correct, and {\bf Scenario 4}: Both models are misspecified. For data generation, we set $\vec\beta=(0,1,1,1,1,1,1,-0.8)^\top$, $\vec\delta=(1,-1,1,1,1,1,1,-0.8,1)^\top$, and $\vec\psi=(1, -2.5, 1.5, 1.5, 1.5, 1.5, 0, 2)^\top$. This parameter setup was chosen because, with this setup under Scenario 4, where both models are misspecified, the coefficient estimate of at least one true effect modifier was estimated to be around zero. This lets us verify the double robustness property of the proposed estimator in terms of effect modifier selection consistency. Data were generated under an exchangeable structure with $\alpha = 0.8$, error variance $\sigma^2_\epsilon=1$ and autocorrelation coefficient $\rho = 0.25$. For each generated data set, we performed the proposed estimation under each of the four scenarios and three working correlation structures. For estimation under Scenario 1, we estimated the propensity score using the true exposure  model but excluded the $\exp(L^{(5)})$ predictor from the treatment-free model. In Scenario 2, we included the $\exp(L^{(5)})$ predictor in the treatment-free model but used a null model for estimating the propensity score, i.e., we assigned the overall empirical proportion of exposure as the probability of being exposed at each measurement occasion. In Scenario 3, we estimated the propensity score using the true exposure  model and also included $\exp(L^{(5)})$ as a covariate in the treatment-free model. Finally, in Scenario 4, we considered a null model for estimating the propensity score and also excluded $\exp(L^{(5)})$ as a covariate in the treatment-free part. We report the percentage selection for each of the candidate effect modifiers with rates of false negative, false positive, and exact selection in Table~\ref{double.robust.tab}. 

\begin{table}[ht]
\centering
\caption{Variable selection rates (\%) from the penalized G-estimation under different scenarios with different working correlation structures. Data were generated with $n=500$, $J=6$ for all $i$, an exchangeable correlation structure with $\alpha=0.8$, error variance $\sigma^2_\epsilon=1$ and autocorrelation coefficient $\rho=0.25$ inducing correlation among the effect modifiers and noise covariates. Results are obtained from 500 independent simulations.}
\resizebox{\textwidth}{!}{
\begin{tabular}{lccccccccccccccc}
  \toprule
  &\multicolumn{3}{c}{\bf Scenario 1}&&\multicolumn{3}{c}{\bf Scenario 2}&&\multicolumn{3}{c}{\bf Scenario 3}&&\multicolumn{3}{c}{\bf Scenario 4}\\
  \cline{2-4}\cline{6-8}\cline{10-12}\cline{14-16}
 & Indep & Exch & UN & & Indep & Exch & UN &  & Indep & Exch& UN&  & Indep & Exch& UN \\ 
  \hline\hline
$A \times L^{(1)}$  & 100 & 100 & 100 &  & 100 & 100 & 100 &  & 100 & 100 & 100 &  & 100 & 100 & 100 \\ 
$A \times L^{(2)}$  & 100 & 100 & 100 &  & 100 & 100 & 100 &  & 100 & 100 & 100 &  & 100 & 100 & 100 \\ 
$A \times L^{(3)}$  & 100 & 100 & 100 &  & 100 & 100 & 100 &  & 100 & 100 & 100 &  & 100 & 100 & 100 \\ 
$A \times L^{(4)}$  & 100 & 100 & 100 &  & 100 & 100 & 100 &  & 100 & 100 & 100 &  & 100 & 100 & 100 \\ 
$A \times L^{(5)}$  & 100 & 100 & 100 &  & 100 & 100 & 100 &  & 100 & 100 & 100 &  & 6.6 & 7.0 & 8.2 \\ 
$A \times L^{(6)}$  & 5.8 & 4.4 & 3.6 &  & 0.0 & 0.0 & 0.0 &  & 2.2 & 0.0 & 0.0 &  & 0.2 & 0.0 & 0.0 \\ 
 $A \times \exp{(L^{(5)})}$  & - & - & - &  & 0.0 & 0.6 & 0.4 &  & 3.2 & 0.0 & 0.0 &  & - & - & - \\ 
 $A \times A_{\text{Lag1}}$ & 100 & 100 & 100 &  & 100 & 100 & 100 &  & 100 & 100 & 100 &  & 100 & 100 & 100 \\ 
 $A \times  X^{(1)}$ & 0.0 & 0.2 & 0.2 &  & 0.2 & 0.0 & 0.0 &  & 2.0 & 0.0 & 0.0 &  & 0.2 & 0.0 & 0.0 \\ 
 $A \times  X^{(2)}$ & 0.4 & 0.2 & 0.0 &  & 0.0 & 0.0 & 0.0 &  & 1.2 & 0.0 & 0.0 &  & 0.0 & 0.0 & 0.0 \\ 
$A \times  X^{(3)}$ & 0.2 & 0.4 & 0.4 &  & 0.2 & 0.0 & 0.0 &  & 2.2 & 0.0 & 0.0 &  & 0.2 & 0.0 & 0.0 \\ 
$A \times  X^{(4)}$ & 0.2 & 0.2 & 0.2 &  & 0.0 & 0.0 & 0.0 &  & 2.8 & 0.0 & 0.0 &  & 0.4 & 0.0 & 0.0 \\ 
$A \times  X^{(5)}$ & 0.0 & 0.2 & 0.4 &  & 0.0 & 0.0 & 0.0 &  & 1.6 & 0.0 & 0.0 &  & 0.0 & 0.0 & 0.0 \\ 
$A \times  X^{(6)}$ & 0.0 & 0.2 & 0.4 &  & 0.0 & 0.0 & 0.0 &  & 2.2 & 0.0 & 0.0 &  & 0.0 & 0.0 & 0.0 \\ 
$A \times  X^{(7)}$ & 0.0 & 0.0 & 0.0 &  & 0.0 & 0.0 & 0.0 &  & 2.2 & 0.0 & 0.0 &  & 0.0 & 0.0 & 0.0 \\ 
$A \times  X^{(8)}$ & 0.4 & 0.4 & 0.4 &  & 0.2 & 0.0 & 0.0 &  & 2.6 & 0.0 & 0.0 &  & 0.0 & 0.0 & 0.0 \\ 
$A \times  X^{(9)}$ & 0.0 & 0.0 & 0.2 &  & 0.0 & 0.0 & 0.0 &  & 3.0 & 0.0 & 0.0 &  & 0.2 & 0.0 & 0.0 \\ 
$A \times  X^{(10)}$ & 0.2 & 0.2 & 0.8 &  & 0.2 & 0.0 & 0.0 &  & 2.8 & 0.0 & 0.0 &  & 0.0 & 0.0 & 0.0 \vspace{0.2cm}\\ 
  FN & 0.0 & 0.0 & 0.0 &  & 0.0 & 0.0 & 0.0 &  & 0.0 & 0.0 & 0.0 &  & 93.4 & 93.0 & 91.8 \\ 
  FP & 6.8 & 6.0 & 5.8 &  & 0.6 & 0.6 & 0.4 &  & 20.6 & 0.0 & 0.0 &  & 1.0 & 0.0 & 0.0 \\ 
  EXACT & 93.2 & 94.0 & 94.2 &  & 99.4 & 99.4 & 99.6 &  & 79.4 & 100 & 100 &  & 6.2 & 7.0 & 8.2 \\ 
  \bottomrule
\end{tabular} \label{double.robust.tab}
}
\footnotesize
FN: \% of false negatives, FP: \% of false positives, EXACT: \% of exact selections, AFP: average false positives,\\ EM: effect modifier, Indep: independent, Exch: exchangeable, UN: unstructured,\\ Scenario 1: treatment model correct, Scenario 2: treatment-free model correct,\\ Scenario 3: both correct, Scenario 4: both incorrect
\end{table}

When both the treatment and the treatment-free models were incorrect (Scenario 4), the proposed method did not select the true effect modifier, $L^{(5)}$, in more than 70\% of the simulations irrespective of what the correlation structure was. Misspecification of the functional form of this effect modifier in the treatment-free part caused a biased estimation of its coefficient in the blip model. The direction of this bias was towards the null, and as a result, we observed the exclusion of this effect modifier in the majority of the simulations. But when at least one of those models was correct (Scenarios 1 to 3) the proposed method performed well in identifying the true effect modifiers under the non-independent correlation structures. When both models were correct (Scenario 3), the false positive rates were higher under the independent structure when compared with exchangeable and unstructured correlation structures. 


\section{Application to the hemodiafiltration study} \label{application}

Our study data arise from an open cohort of patients undergoing chronic hemodiafiltration at CHUM and CED. Hemodiafiltration was considered as chronic if there were least 28 consecutive sessions. The cohort start-date for each patient was their first dialysis session on or after  March 1st, 2017; the cohort end date was December 1st, 2021. 
The primary data include information from a total of 474 patients who underwent  170761  dialysis sessions. The following information was  extracted from hospital databases for each session: drugs (BDM), laboratory results (CERNER), procedures related to dialysis venous access (Radimage), and dialysis-related variables and dialysis-related drugs (EuCliD-NephroCare). 
Comorbidities that are potential confounders were obtained from the \emph{Maintenance et exploitation des données pour l'étude de la clientèle hospitalière} (MED-ECHO) database. Confounders and potential effect modifiers that we considered in the analysis are previous outcome (24L or less$= 1$ vs.\ more than 24L $= 0$), hemoglobin, albumin, dalteparin, access (fistula = 0, catheter = 1), catheter change, age, sex, and the components of the Charlson Comorbidity Index (hypertension, diabetes, peripheral vascular disease, congestive heart failure, cardiac arrhythmia, acute myocardial infarction, chronic pulmonary disease, liver disease, valvular disease, cancer, metastatic cancer, cerebrovascular disease, dementia, 
hemiplegia, and rheumatic disease). The exposure is the dialysis facility and coded as one if the treatment location was CHUM and zero if the location was CED. Some descriptive statistics of the data are presented in Section~\ref{hdf.descriptives} of the Appendix.

We included the first six consecutive sessions of post-dilution hemodiafiltration for each of the patients in our analysis. We estimated the propensity scores using a logistic regression of exposure conditional on all of the potential confounders using the pooled data set. We performed the proposed penalized G-estimation considering four different correlation structures: independent, exchangeable, autoregressive of order one (AR1), and unstructured. For $j$-th session, for example, under the AR1 correlation structure, the selected blip model (with adjustment for all potential confounders in the treatment-free part) is 
\begin{align*}
    \gamma_j(a_{j}, \vec h_{j};\vec\psi)
    = (\psi_0 + \psi_1\times\text{Cancer}_j)\,\text{CHUM}_j
\end{align*}
for $j=1,2,\ldots,6$. 
The estimates of the blip parameters are given in Table~\ref{tab:application} with their corresponding sandwich standard error estimates.

\begin{table}[ht]
\centering
\caption{Estimated blip parameters and corresponding standard errors under different working correlation structures for the hemodiafiltration study.}
\begin{tabular}{lccccccccccc}
  \toprule
  &\multicolumn{2}{c}{\bf Indep}&&\multicolumn{2}{c}{\bf Exch}&&\multicolumn{2}{c}{\bf AR1}&&\multicolumn{2}{c}{\bf UN}\\
  \cline{2-3}\cline{5-6}\cline{8-9}\cline{11-12}
 &Est&SE&&Est&SE&&Est&SE&&Est&SE \\ 
  \hline\hline
CHUM & -0.83 & 0.24 &  & -1.51 & 0.32 &  & -1.85 & 0.31 &  & -1.71 & 0.34 \\ 
CHUM$\times$cancer & - & - &  & - & - &  & 3.89 & 0.78 &  & 3.37 & 0.75 \\ 
   \hline
\end{tabular} \label{tab:application}
\end{table}

The sign of the estimated main effect of the exposure is negative under each of the working correlation structures considered. Under the AR1 and unstructured correlation structures,  cancer was selected by the method, indicating that the effect of dialysis facility on the convection volume differs by the cancer status of the patient. If we interpret the results obtained under the AR1 correlation structure we can say that among the patients who did not have cancer the mean convection volume was 1.85 litres lower at CHUM when compared with CED at fixed levels of all other confounders. But, the mean convection volume was $3.89-1.85=2.04$ litres higher at the CHUM for cancer patients indicating that cancer patients with same measurements on the adjusted confounders had better hemodiafiltration outcomes at the CHUM. However, as highlighted in Remark 2 of Section~\ref{sec.asymp}, it is important to note that the final model was determined by data-driven variable selection. So confidence intervals based on the sandwich standard error estimates may produce invalid inference with inflated type I error. 

\section{Discussion} \label{discussion}

In this paper, we proposed a penalization in the G-estimation for evaluating the causal effect of a time-varying exposure with automatic effect modifier selection with longitudinal observational data and a proximal outcome of interest. We applied this method to investigate if the effect of dialysis facility on the dialysis outcome (convection volume) differs by the demographic and clinical characteristics and comorbidity status of the patients with end-stage renal disease. Our findings suggest that while the CED produced better hemodiafiltration outcomes overall, cancer patients with similar measured characteristics might have had better outcomes at the CHUM compared to the CED.


In the simulation study, our method performed well in selecting the true blip model from a set of candidate models, even when we assumed no correlation between the outcomes from the same patient. However, when we used the correct correlation structure, the method produced better estimates of the target (blip) parameters than a working independence model. 
Although we considered different non-independent correlation structures in the simulation studies and in the data application, 
it is possible to choose an appropriate correlation structure using the data \citep{jaman2016determinant, inan2019press, sultana2023caution}. The selection rates of the true blip model were good in the majority of our simulation scenarios. However, these rates  depend on the magnitude of effect modification and the signal to noise ratio.  A challenge still remains in the identification of weak effect modifiers and this would be an interesting topic for future research. Researchers may wish to study the article by \cite{gunter2011variable}, where the authors discussed a ranking procedure for selecting effect modifiers that are weaker but important for decision making.

Other limitations of our method include the computational burden that may arise when subjects contribute a large number of observations. We also did not address informative censoring in this paper, but this is a possibility for a future extension. Although we did not perform variable selection for the treatment model in our hemodiafiltarion application, such selection may be necessary in practice, especially in a high-dimensional setup (see the work by \cite{shortreed2017outcome} for related method).\\ 

{\bf \large Software implementation}\\
The R-codes for implementing our method are available at (\url{https://github.com/ajmeryjaman/penalizedG}).\\

{\bf \large Funding}\\
This work is supported by a doctoral scholarship from the Fonds de Recherche du Québec Nature et technologies (FRQNT) of Canada to AJ and a Discovery Grant from the Natural Sciences and Engineering Research Council of Canada to MES. AE is supported by the research grants (R01DA058996, R01DA048764, and R33NS120240) from the National Institutes of Health. MES is supported by a tier 2 Canada Research Chair.
\\

\bibliographystyle{apalike}
\bibliography{bibliography.bib}

\begin{thebibliography}{}

\bibitem[Ashley, 2015]{ashley2015precision}
Ashley, E.~A. (2015).
\newblock The precision medicine initiative: a new national effort.
\newblock {\em Jama}, 313(21):2119--2120.

\bibitem[Bahamyirou et~al., 2022]{bahamyirou2022doubly}
Bahamyirou, A., Schnitzer, M.~E., Kennedy, E.~H., Blais, L., and Yang, Y. (2022).
\newblock Doubly robust adaptive lasso for effect modifier discovery.
\newblock {\em The International Journal of Biostatistics}.

\bibitem[Balan and Schiopu-Kratina, 2005]{balan2005asymptotic}
Balan, R. and Schiopu-Kratina, I. (2005).
\newblock Asymptotic results with generalized estimating equations for longitudinal data.
\newblock {\em Ann. Statist.}, 33(1):522--541.

\bibitem[Bian et~al., 2021]{bian2021variable}
Bian, Z., Moodie, E.~E., Shortreed, S.~M., and Bhatnagar, S. (2021).
\newblock Variable selection in regression-based estimation of dynamic treatment regimes.
\newblock {\em Biometrics}.

\bibitem[Bian et~al., 2023]{bian2023variable}
Bian, Z., Moodie, E.~E., Shortreed, S.~M., Lambert, S.~D., and Bhatnagar, S. (2023).
\newblock Variable selection for individualised treatment rules with discrete outcomes.
\newblock {\em Journal of the royal statistical society. Series C (Applied statistics)}, qlad096, https://doi.org/10.1093/jrsssc/qlad096.

\bibitem[Boruvka et~al., 2018]{boruvka2018assessing}
Boruvka, A., Almirall, D., Witkiewitz, K., and Murphy, S.~A. (2018).
\newblock Assessing time-varying causal effect moderation in mobile health.
\newblock {\em Journal of the American Statistical Association}, 113(523):1112--1121.

\bibitem[Chakraborty and Moodie, 2013]{chakraborty2013statistical}
Chakraborty, B. and Moodie, E.~E. (2013).
\newblock Statistical methods for dynamic treatment regimes.
\newblock {\em Springer-Verlag. doi}, 10(978-1):4--1.

\bibitem[Chakraborty and Murphy, 2014]{chakraborty2014dynamic}
Chakraborty, B. and Murphy, S.~A. (2014).
\newblock Dynamic treatment regimes.
\newblock {\em Annual review of statistics and its application}, 1:447--464.

\bibitem[Chapdelaine et~al., 2015]{chapdelaine2015optimization}
Chapdelaine, I., de~Roij~van Zuijdewijn, C.~L., Mostovaya, I.~M., L{\'e}vesque, R., Davenport, A., Blankestijn, P.~J., Wanner, C., Nub{\'e}, M.~J., Grooteman, M.~P., Group, E., et~al. (2015).
\newblock Optimization of the convection volume in online post-dilution haemodiafiltration: practical and technical issues.
\newblock {\em Clinical kidney journal}, 8(2):191--198.

\bibitem[Fan and Li, 2001]{fan2001variable}
Fan, J. and Li, R. (2001).
\newblock Variable selection via nonconcave penalized likelihood and its oracle properties.
\newblock {\em Journal of the American statistical Association}, 96(456):1348--1360.

\bibitem[Fan and Tang, 2013]{fan2013tuning}
Fan, Y. and Tang, C.~Y. (2013).
\newblock Tuning parameter selection in high dimensional penalized likelihood.
\newblock {\em Journal of the Royal Statistical Society Series B: Statistical Methodology}, 75(3):531--552.

\bibitem[Grooteman et~al., 2012]{grooteman2012effect}
Grooteman, M.~P., van~den Dorpel, M.~A., Bots, M.~L., Penne, E.~L., van~der Weerd, N.~C., Mazairac, A.~H., den Hoedt, C.~H., van~der Tweel, I., L{\'e}vesque, R., Nub{\'e}, M.~J., et~al. (2012).
\newblock Effect of online hemodiafiltration on all-cause mortality and cardiovascular outcomes.
\newblock {\em Journal of the American Society of Nephrology: JASN}, 23(6):1087.

\bibitem[Gunter et~al., 2011]{gunter2011variable}
Gunter, L., Zhu, J., and Murphy, S. (2011).
\newblock Variable selection for qualitative interactions.
\newblock {\em Statistical methodology}, 8(1):42--55.

\bibitem[He et~al., 2015]{he2015structural}
He, J., Stephens-Shields, A., and Joffe, M. (2015).
\newblock Structural nested mean models to estimate the effects of time-varying treatments on clustered outcomes.
\newblock {\em The International Journal of Biostatistics}, 11(2):203--222.

\bibitem[Hunter and Li, 2005]{hunter2005variable}
Hunter, D.~R. and Li, R. (2005).
\newblock Variable selection using mm algorithms.
\newblock {\em Annals of statistics}, 33(4):1617.

\bibitem[Inan et~al., 2019]{inan2019press}
Inan, G., Latif, M.~A., and Preisser, J. (2019).
\newblock A press statistic for working correlation structure selection in generalized estimating equations.
\newblock {\em Journal of Applied Statistics}, 46(4):621--637.

\bibitem[Jaman et~al., 2016]{jaman2016determinant}
Jaman, A., Latif, M.~A., Bari, W., and Wahed, A.~S. (2016).
\newblock A determinant-based criterion for working correlation structure selection in generalized estimating equations.
\newblock {\em Statistics in Medicine}, 35(11):1819--1833.

\bibitem[Johnson et~al., 2008]{johnson2008penalized}
Johnson, B.~A., Lin, D., and Zeng, D. (2008).
\newblock Penalized estimating functions and variable selection in semiparametric regression models.
\newblock {\em Journal of the American Statistical Association}, 103(482):672--680.

\bibitem[Jones et~al., 2022]{jones2022valid}
Jones, J., Ertefaie, A., and Strawderman, R.~L. (2022).
\newblock Valid post-selection inference in robust q-learning.
\newblock {\em arXiv preprint arXiv:2208.03233}.

\bibitem[Koch et~al., 2018]{koch2018covariate}
Koch, B., Vock, D.~M., and Wolfson, J. (2018).
\newblock Covariate selection with group lasso and doubly robust estimation of causal effects.
\newblock {\em Biometrics}, 74(1):8--17.

\bibitem[Liang and Zeger, 1986]{liang1986longitudinal}
Liang, K.-Y. and Zeger, S.~L. (1986).
\newblock Longitudinal data analysis using generalized linear models.
\newblock {\em Biometrika}, 73(1):13--22.

\bibitem[Ma et~al., 2023]{ma2023sequential}
Ma, T., Cai, H., Qi, Z., Shi, C., and Laber, E.~B. (2023).
\newblock Sequential knockoffs for variable selection in reinforcement learning.
\newblock {\em arXiv preprint arXiv:2303.14281}.

\bibitem[Marcelli et~al., 2015]{marcelli2015high}
Marcelli, D., Scholz, C., Ponce, P., Sousa, T., Kopperschmidt, P., Grassmann, A., Pinto, B., and Canaud, B. (2015).
\newblock High-volume postdilution hemodiafiltration is a feasible option in routine clinical practice.
\newblock {\em Artificial organs}, 39(2):142--149.

\bibitem[Moodie et~al., 2023]{moodie2023variable}
Moodie, E.~E., Bian, Z., Coulombe, J., Lian, Y., Yang, A.~Y., and Shortreed, S.~M. (2023).
\newblock Variable selection in high dimensions for discrete-outcome individualized treatment rules: Reducing severity of depression symptoms.
\newblock {\em Biostatistics}, page kxad022.

\bibitem[Murphy, 2003]{murphy2003optimal}
Murphy, S.~A. (2003).
\newblock Optimal dynamic treatment regimes.
\newblock {\em Journal of the Royal Statistical Society Series B: Statistical Methodology}, 65(2):331--355.

\bibitem[Pan, 2001]{pan2001akaike}
Pan, W. (2001).
\newblock Akaike's information criterion in generalized estimating equations.
\newblock {\em Biometrics}, 57(1):120--125.

\bibitem[Peters et~al., 2016]{peters2016haemodiafiltration}
Peters, S.~A., Bots, M.~L., Canaud, B., Davenport, A., Grooteman, M.~P., Kircelli, F., Locatelli, F., Maduell, F., Morena, M., Nub{\'e}, M.~J., et~al. (2016).
\newblock Haemodiafiltration and mortality in end-stage kidney disease patients: a pooled individual participant data analysis from four randomized controlled trials.
\newblock {\em Nephrology Dialysis Transplantation}, 31(6):978--984.

\bibitem[Robins, 1992]{robins1992estimation}
Robins, J. (1992).
\newblock Estimation of the time-dependent accelerated failure time model in the presence of confounding factors.
\newblock {\em Biometrika}, 79(2):321--334.

\bibitem[Robins and Hernan, 2008]{robins2008estimation}
Robins, J. and Hernan, M. (2008).
\newblock Estimation of the causal effects of time-varying exposures.
\newblock {\em Chapman \& Hall/CRC Handbooks of Modern Statistical Methods}, pages 553--599.

\bibitem[Robins, 1989]{robins1989analysis}
Robins, J.~M. (1989).
\newblock The analysis of randomized and non-randomized aids treatment trials using a new approach to causal inference in longitudinal studies.
\newblock {\em Health service research methodology: a focus on AIDS}, pages 113--159.

\bibitem[Robins, 1997]{robins1997causal}
Robins, J.~M. (1997).
\newblock Causal inference from complex longitudinal data.
\newblock In {\em Latent variable modeling and applications to causality}, pages 69--117. Springer.

\bibitem[Robins et~al., 2007]{robins2007invited}
Robins, J.~M., Hernan, M.~A., and Rotnitzky, A. (2007).
\newblock Invited commentary: effect modification by time-varying covariates.
\newblock {\em American Journal of Epidemiology}, 166(9):994--1002.

\bibitem[Ronco and Cruz, 2007]{ronco2007hemodiafiltration}
Ronco, C. and Cruz, D. (2007).
\newblock Hemodiafiltration history, technology, and clinical results.
\newblock {\em Advances in Chronic Kidney Disease}, 14(3):231--243.

\bibitem[Shi et~al., 2018]{shi2018high}
Shi, C., Fan, A., Song, R., and Lu, W. (2018).
\newblock High-dimensional a-learning for optimal dynamic treatment regimes.
\newblock {\em Annals of statistics}, 46(3):925.

\bibitem[Shortreed and Ertefaie, 2017]{shortreed2017outcome}
Shortreed, S.~M. and Ertefaie, A. (2017).
\newblock Outcome-adaptive lasso: variable selection for causal inference.
\newblock {\em Biometrics}, 73(4):1111--1122.

\bibitem[Sultana et~al., 2023]{sultana2023caution}
Sultana, A., Lipi, N., and Jaman, A. (2023).
\newblock A caution in the use of multiple criteria for selecting working correlation structure in generalized estimating equations.
\newblock {\em Communications in Statistics-Simulation and Computation}, 52(3):980--992.

\bibitem[Tang et~al., 2023]{tang2023ultra}
Tang, D., Kong, D., Pan, W., and Wang, L. (2023).
\newblock Ultra-high dimensional variable selection for doubly robust causal inference.
\newblock {\em Biometrics}, 79(2):903--914.

\bibitem[Tibshirani, 1996]{tibshirani1996regression}
Tibshirani, R. (1996).
\newblock Regression shrinkage and selection via the lasso.
\newblock {\em Journal of the Royal Statistical Society: Series B (Methodological)}, 58(1):267--288.

\bibitem[Tsiatis, 2006]{tsiatis2006semiparametric}
Tsiatis, A.~A. (2006).
\newblock {\em Semiparametric theory and missing data}.
\newblock Springer.

\bibitem[VanderWeele and Robins, 2007]{vanderweele2007four}
VanderWeele, T.~J. and Robins, J.~M. (2007).
\newblock Four types of effect modification: a classification based on directed acyclic graphs.
\newblock {\em Epidemiology (Cambridge, Mass.)}, 18(5):561--568.

\bibitem[Vansteelandt and Joffe, 2014]{vansteelandt2014structural}
Vansteelandt, S. and Joffe, M. (2014).
\newblock Structural nested models and g-estimation: The partially realized promise.
\newblock {\em Statistical Science}, 29(4):707--731.

\bibitem[Wahba, 1985]{wahba1985comparison}
Wahba, G. (1985).
\newblock A comparison of gcv and gml for choosing the smoothing parameter in the generalized spline smoothing problem.
\newblock {\em The annals of statistics}, pages 1378--1402.

\bibitem[Wallace and Moodie, 2015]{wallace2015doubly}
Wallace, M.~P. and Moodie, E.~E. (2015).
\newblock Doubly-robust dynamic treatment regimen estimation via weighted least squares.
\newblock {\em Biometrics}, 71(3):636--644.

\bibitem[Wallace et~al., 2019]{wallace2019model}
Wallace, M.~P., Moodie, E.~E., and Stephens, D.~A. (2019).
\newblock Model selection for g-estimation of dynamic treatment regimes.
\newblock {\em Biometrics}, 75(4):1205--1215.

\bibitem[Wang et~al., 2009]{wang2009shrinkage}
Wang, H., Li, B., and Leng, C. (2009).
\newblock Shrinkage tuning parameter selection with a diverging number of parameters.
\newblock {\em Journal of the Royal Statistical Society Series B: Statistical Methodology}, 71(3):671--683.

\bibitem[Wang et~al., 2007]{wang2007tuning}
Wang, H., Li, R., and Tsai, C.-L. (2007).
\newblock Tuning parameter selectors for the smoothly clipped absolute deviation method.
\newblock {\em Biometrika}, 94(3):553--568.

\bibitem[Wang et~al., 2012]{wang2012penalized}
Wang, L., Zhou, J., and Qu, A. (2012).
\newblock Penalized generalized estimating equations for high-dimensional longitudinal data analysis.
\newblock {\em Biometrics}, 68(2):353--360.

\bibitem[Zhao et~al., 2022]{zhao2022selective}
Zhao, Q., Small, D.~S., Ertefaie, A., et~al. (2022).
\newblock Selective inference for effect modification via the lasso.
\newblock {\em Journal of the Royal Statistical Society Series B}, 84(2):382--413.

\end{thebibliography}

\appendix

\section{Appendix}
\subsection{Form of efficient score} \label{form}
The efficient score vector is $\vec S^{\text{eff}}(\vec\theta)=\{\vec S^{\text{eff}}(\vec\delta)^\top, \vec S^{\text{eff}}(\vec\psi)^\top\}^\top$, where
\begin{align*}
    \vec S^{\text{eff}}(\vec\delta)&=\sum_{i=1}^n \Big\{\frac{\partial\vec\mu(\vec H_i;\vec\delta)}{\partial\vec\delta^\top}\Big\}^\top\,Var(\vec U_i|\vec H_i)^{-1}\{\vec U_i - E(\vec U_i|\vec H_i)\}\\
    &=\sum_{i=1}^n \vec H_i^\top\,Var(\vec U_i|\vec H_i)^{-1}\{\vec U_i - E(\vec U_i|\vec H_i)\}\\
    \vec S^{\text{eff}}(\vec\psi)&=\sum_{i=1}^n \Big[\frac{\partial\vec\gamma^*(\vec A_i,\vec H_i;\vec\psi)}{\partial\vec\psi^\top} - E\Big\{\frac{\partial\vec\gamma^*(\vec A_i,\vec H_i;\vec\psi)}{\partial\vec\psi^\top}|\vec H_i\Big\}\Big]^\top\,Var(\vec U_i|\vec H_i)^{-1}\{\vec U_i - E(\vec U_i|\vec H_i)\}\\
    &=\sum_{i=1}^n \Big[\{\vec A_i-E(\vec A_i|\vec H_i)\}\vec H_i\Big]^\top\,Var(\vec U_i|\vec H_i)^{-1}\{\vec U_i - E(\vec U_i|\vec H_i)\}
\end{align*}


\subsection{Method of moments estimators for $\sigma^2$ and $\alpha$} \label{moment.method}

Calculate the variance parameter as
\begin{align*}
    \hat{\sigma}^2 =  \frac{1}{n}\sum_{i=1}^n\frac{1}{J}\sum_{j=1}^{J} e_{ij}^2
\end{align*}
and the correlation parameter(s) as shown in Table~\ref{moment.method.est}.
\begin{table}[ht]
    \centering
    \caption{Common choices for working correlation structures and corresponding estimators of the correlation parameter.}
    \begin{tabular}{ccc} 
    \toprule
   Working structure & $\widehat{\text{Corr}}(U_{ij},U_{ik})$ & Estimator \\
    \hline
   Independence &0 & -\\
   Exhangeable & $\hat{\alpha}$ & $\hat{\alpha}=\frac{1}{n\hat{\sigma}^2}\sum_{i=1}^n\frac{1}{J(J-1)}\underset{j\neq k}{\sum\sum} e_{ij}e_{ik}$\\
    AR(1)&$\hat{\alpha}^{|j-k|}$ &$\hat{\alpha}=\frac{1}{n\hat{\sigma}^2}\sum_{i=1}^n\frac{1}{J-1}\underset{j\leq J-1}{\sum} e_{ij}e_{i,(j+1)}$\\
    Unstructured&$\hat{\alpha}_{jk}$ &$\hat{\alpha}_{jk}=\frac{1}{n\hat{\sigma}^2}\sum_{i=1}^n e_{ij}e_{ik}$\\
    \bottomrule
    \end{tabular}
    \label{moment.method.est}
\end{table}


\subsection{Tuning parameter selection}\label{tune.par.selection}

The oracle properties of the penalized estimators rely on the proper choice of tuning parameter. \cite{johnson2008penalized} considered the generalized cross-vaidation (GCV) statistic \citep{wahba1985comparison} for selecting the tuning parameter in penalized estimating equations. The use of GCV for such selection was suggested by \cite{tibshirani1996regression} and \cite{fan2001variable}. However, \cite{wang2007tuning} showed that tuning parameter selection using GCV causes a nonignorable overfitting effect even if the sample size goes to infinity. Considering the number of covariates $K$ as fixed, \cite{wang2007tuning} proposed the Bayesian information criterion (BIC) for tuning parameter selection which showed consistency in the identification of the true model. 
Later, a modification in the BIC criterion was proposed by \cite{wang2009shrinkage} in a moderately high-dimensional setup ($K < n$) and by \cite{fan2013tuning} in an ultra-high dimensional setup ($K >> n$). All of these criteria have the following general form
\begin{align} \label{general.criterion}
    \text{measure of model fit} + \tau_n \times \text{measure of model complexity},
\end{align}
where $\tau_n$ is a positive sequence.

In the context of variable selection for individualized treatment rules, \cite{bian2023variable} and  \cite{moodie2023variable} used doubly robust information criteria for tuning parameter selection. These criteria are similar to the criterion of \cite{fan2013tuning}, except that \cite{bian2023variable} and \cite{moodie2023variable} introduced a weighted measure of model fitting that incorporates overlap weights. To construct the measure of model fitting we follow an approach similar to \cite{wang2007tuning} and construct the weighted loss function following \cite{moodie2023variable}. We define the weighted loss function corresponding to the G-estimating equations as
\begin{align}
    L(\tilde{\vec\theta}_{\lambda_n}) = \sum_{i=1}^n\sum_{j=1}^J |\vec A_{ij}-E(\vec A_{ij}|\vec H_{ij})|\times (\vec Y_{ij} - \vec H_{ij}^\top\tilde{\vec\delta}_{\lambda_n} - \vec A_{ij}\vec H_{ij}^\top\tilde{\vec\psi}_{\lambda_n})^2,
\end{align}
where $|\vec A_{ij}-E(\vec A_{ij}|\vec H_{ij})|$ are the overlap weights and $\tilde{\vec\theta}_{\lambda_n}$ denotes the penalized estimate obtained under a fixed value of the tuning parameter $\lambda_n$. Since the joint distribution of the data under a non-diagonal correlation structure is unknown/undefined, we construct the loss function associated with the estimating equations under an independence assumption. For the penalized estimating equations, the measure of model complexity can be defined with the following generalized degrees of freedom \citep{johnson2008penalized}
\begin{align*}
    \text{DF}_{\lambda_n} = trace\Bigg[\Big\{\hat{\vec H}_n(\tilde{\vec\theta}_{\lambda_n})+n\vec E_n(\tilde{\vec\theta}_{\lambda_n})\Big\}^{-1}\hat{\vec H}_n(\tilde{\vec\theta}_{\lambda_n})\Bigg].
\end{align*}
In our setup, we propose the following doubly robust information criterion:
\begin{align} \label{dric}
    \text{DRIC}_{\lambda_n} &= \log\Bigg\{\frac{L(\tilde{\vec\theta}_{\lambda_n})}{n\times J}\Bigg\} + \tau_n\times\dfrac{\text{DF}_{\lambda_n}}{n}.
\end{align}
We set $\tau_n=\log\{\log(n)\}\log(2K)$, a consideration similar to \cite{fan2013tuning}, where $2K$ represents the dimension of $\vec\theta$. Note that we can even consider $\tau_n=\log(n)$ in the low dimensional setting.

We compute the DRIC for a sequence of values for $\lambda_n$ which are in decreasing order. Typically, the first element ($\lambda_{\text{max}}$) in this sequence is the lowest positive number such that all the effect modifiers are eliminated by the method. The last element ($\lambda_{\text{min}}$) is a value close to zero for which none of the effect modifiers are eliminated. And we consider around one hundred values in between so that we have a fine grid. The optimal tuning parameter is the value of $\lambda_n$ that corresponds to the minimum value of $\text{DRIC}_{\lambda_n}$.

Note that \cite{wang2012penalized} used cross-validation for selecting the tuning parameter in a penalized generalized estimating equation, minimizing the prediction error under an independence assumption. We followed a similar idea and constructed the loss function in the DRIC using an independent correlation structure. Although we considered a repeated outcomes setup, in our study, we did not consider the quasi-information criterion (QIC) \citep{pan2001akaike}, which extends the AIC in the longitudinal setup. Because the QIC might experience similar overfitting issues as the AIC and since our goal is to select the true effect modifiers in SNMMs, we instead used the DRIC for tuning parameter selection. Structurally, DRIC has resemblance to the BIC.

\subsection{Regularity conditions} \label{regularity.cond}
For establishing the asymptotic theory of the penalized G-estimator, we  need the following regularity conditions \citep{wang2012penalized} to hold, some of which maybe further relaxed. Majority of these conditions are analogous to the regularity conditions for the generalized estimating equations for longitudinal data.
\begin{enumerate}
    \item [(C1)] All variables in $\vec D_{ij}$, $i=1,\ldots,n$, $j=1,\ldots,J$, are uniformly bounded.
    \item [(C2)] The unknown parameter $\vec\theta_n$ belongs to a compact subset $\vec\Theta \subseteq R^{2K}$ and the true parameter $\vec\theta_0$ lies in the interior of $\vec\Theta$.
    \item [(C3)] There exists finite positive constants $c_1$ and $c_2$ such that
    \begin{align*}
        c_1 \leq \omega_{min} \Bigg(\dfrac{\sum_{i=1}^n\vec D_i^\top(\vec H_i\;\;\vec A_i\cdot\vec H_i)}{n}\Bigg) \leq \omega_{max} \Bigg(\dfrac{\sum_{i=1}^n\vec D_i^\top(\vec H_i\;\;\vec A_i\cdot\vec H_i)}{n}\Bigg)\leq c_2,
    \end{align*}
    where $\omega_{min}(\vec D)$ and $\omega_{max}(\vec D)$ denote the minimum and maximum of the eigenvalues, respectively, of the matrix $\vec D$.
    \item [(C4)] The common true correlation matrix $\vec R_0$ for the observed outcomes has eigen values bounded away from zero and $+\infty$. The estimated working correlation matrix $\hat{\vec R}$ satisfies $||\hat{\vec R}^{-1} - \overline{\vec R}^{-1}|| = O_p(\sqrt{1/n})$, 
    where $\overline{\vec R}$ is a constant positive definite matrix with eigen values bounded away from zero and $+\infty$. 
    Note that $||\vec D||=\{trace(\vec D\vec D^\top)\}^{1/2}$ denotes the Frobenius norm of the matrix $\vec D$.
    \item [(C5)] Let $\vec\xi_i(\vec\theta_n) = (\vec\xi_{i1}(\vec\theta_n),\ldots,\vec\xi_{in_i}(\vec\theta_n))^\top=\vec Q_i^{-1/2}(\vec Y_i - \vec g_i(\vec\theta_n))$. There exists a finite constant $d_1 > 0$ such that $E(||\vec\xi_i(\vec\theta_0)||^{2+\kappa}) \leq d_1$ for all $i$ and some $\kappa > 0$; and there esists positive constants $d_2$ and $d_3$ such that $E(\exp(d_2|\xi_{ij}(\vec\theta_0))|\vec D_i) \leq d_3$, uniformly in $i=1,\ldots,n$, $j=1,\ldots,J$. 
    \item [(C6)] Let $T_n=\{\vec\theta_n: ||\vec\theta_n - \vec\theta_0||\leq \Delta\sqrt{1/n}\}$, then $g'(\vec D_{ij}\vec\theta_n)$, $i=1,\ldots,n$, $j=1,\ldots,J$, are uniformly bounded away from 0 and $\infty$ on $T_n$; $g''(\vec D_{ij}\vec\theta_n)$ and $g'''(\vec D_{ij}\vec\theta_n)$, $i=1,\ldots,n$, $j=1,\ldots,J$, are uniformly bounded by a finite positive constant $d_2$ on $T_n$; $g'(.)$, $g''(.)$ and $g'''(.)$ denote the first, second and third derivatives of the function $g(.)$, respectively.
    \item [(C7)] When $s$ is not fixed, assuming $\min_{m \in B} |\theta_{0m}|/\lambda_n \rightarrow \infty$ as $n \rightarrow \infty$ and $s_n^3n^{-1} = o(1)$, $\lambda_n \rightarrow 0$, $s_n^2(\log n)^4 = o(n\lambda_n^2)$, $\log(K_n) = o(n\lambda_n^2/(\log n)^2)$, $K_ns_n^4(\log n)^6 = o(n^2\lambda_n^2)$, and $K_ns_n^3(\log n)^8 = o(n^2\lambda_n^4)$. Note that $\lambda_n$ is the tuning parameter.
\end{enumerate}

\subsection{Proof of Theorem 1} \label{proof.thm1}

 We consider the sets $F_m=\{\tilde{\theta}_m\neq0\}$ such that $m \in B^c$. To prove this theorem it is sufficient to show that for any $\epsilon > 0$, $P(F_m) < \epsilon$ as $n \rightarrow \infty$. Since $\tilde{\theta}_m=O_p(n^{-1/2})$, there exists some $Z$ such that
\begin{align} \label{T11}
    P(F_m) < \epsilon/2 + P(\tilde{\theta}_m\neq 0, |\tilde{\theta}_m| < Zn^{-1/2}) \text{ as } n\rightarrow \infty.
\end{align}
If $\tilde{\vec\theta}$ is an approximate zero crossing, on the set $\{\tilde{\theta}_m\neq 0, |\tilde{\theta}_m| < Zn^{-1/2}\}$ we have,
\begin{align*}
    \{n^{-1/2}S^{\text{eff}}_m(\theta_0)+n^{1/2}\vec H_{(m)}(\vec\theta_0)(\tilde{\theta}-\theta_0) + o_p(1) - n^{1/2}q_{\lambda_n}(|\tilde{\theta}_m|)\text{sign}(\tilde{\theta}_m)\}^2 = o_p(1),
\end{align*}
where $\vec H_{(m)}(\vec\theta_0)$ is the $m$-th row of $\vec H(\vec\theta_0)$. In the above equation, the first three terms are of order $O_p(1)$. Hence, there exists some $Z'$ such that
\begin{align} \label{T12}
    P(\tilde{\theta}_m\neq 0, |\tilde{\theta}_m| < Zn^{-1/2},n^{1/2}q_{\lambda_n}(|\tilde{\theta}_m|) > Z') < \epsilon/2 \text{ as } n\rightarrow \infty.
\end{align}
If the condition (C.b.2) holds, $\tilde{\theta}_m\neq 0$ and $|\tilde{\theta}_m| < Zn^{-1/2}$ together imply that $n^{1/2}q_{\lambda_n}(|\tilde{\theta}_m|) > Z'$ as $n\rightarrow \infty$. Therefore, $P(\tilde{\theta}_m\neq 0, |\tilde{\theta}_m| < Zn^{-1/2}) = P(\tilde{\theta}_m\neq 0, |\tilde{\theta}_m| < Zn^{-1/2},n^{1/2}q_{\lambda_n}(|\tilde{\theta}_m|) > Z')$ and from equations (\ref{T11}) and (\ref{T12}) we obtain that
\begin{align}
    P(F_m) < \epsilon/2 + P(\tilde{\theta}_m\neq 0, |\tilde{\theta}_m| < Zn^{-1/2},n^{1/2}q_{\lambda_n}(|\tilde{\theta}_m|) > Z') < \epsilon.
\end{align}   

\subsection{Proof of Theorem 2} \label{proof.thm2}

  Let $\vec S^{\text{eff}}_B(\vec\theta)$ represents the vector of elements in $\vec S^{\text{eff}}(\vec\theta)$ corresponding to $\theta_m$'s such that $m \in B$ and $\vec H_{B}(\vec\theta)$  represents the $s\times s$  submatrice of  $\vec H(\vec\theta)$. Under conditions (C.a) and (C.b.1)
\begin{align} \label{T2.1}
    n^{-1/2}\vec S^{\text{eff}}_B(\vec\theta_0)+n^{1/2}\vec H_{B}(\vec\theta_0)(\tilde{\vec\theta}_n^B-\vec\theta^B_0) -n^{1/2}\vec q_{\lambda_n}(|\tilde{\vec\theta}_n^B|)\text{sign}(\tilde{\vec\theta}_n^B) = \vec o_p(1).
\end{align}
If we apply the Taylor series approximation of $\vec q_{\lambda_n}(|\tilde{\vec\theta}_n^B|)\text{sign}(\tilde{\vec\theta}_n^B)$ around $\vec\theta^B_0$ in equation (\ref{T2.1}), we obtain
\begin{align*}
    \vec o_p(1) &\approx n^{-1/2}\vec S^{\text{eff}}_B(\vec\theta_0)+n^{1/2}\vec H_{B}(\vec\theta_0)(\tilde{\vec\theta}_n^B-\vec\theta^B_0) - n^{1/2}\Big\{\vec q_{\lambda_n}(|\vec\theta^B_0|)\text{sign}(\vec\theta^B_0)+\vec q'_{\lambda_n}(|\vec\theta^B_0|)\text{sign}(\vec\theta^B_0)(\tilde{\vec\theta}_n^B-\vec\theta^B_0)\Big\}\\
    &=n^{-1/2}\vec S^{\text{eff}}_B(\vec\theta_0)+n^{1/2}\Big\{\vec H_{B}(\vec\theta_0)+\vec W_n(\vec\theta^B_0)\Big\}(\tilde{\vec\theta}_n^B-\vec\theta^B_0)-n^{1/2}\vec q_{\lambda_n}(|\vec\theta^B_0|)\text{sign}(\vec\theta^B_0)\\
    &=n^{-1/2}\vec S^{\text{eff}}_B(\vec\theta_0)+n^{1/2}\Big\{\vec H_{B}(\vec\theta_0)+\vec W_n(\vec\theta^B_0)\Big\}\Big[\tilde{\vec\theta}_n^B-\vec\theta^B_0-\Big\{\vec H_{B}(\vec\theta_0)+\vec W_n(\vec\theta^B_0)\Big\}^{-1}\vec q_{\lambda_n}(|\vec\theta^B_0|)\text{sign}(\vec\theta^B_0)\Big]\\
    &=n^{-1/2}\vec S^{\text{eff}}_B(\vec\theta_0)+n^{1/2}\Big\{\vec H_{B}(\vec\theta_0)+\vec W_n(\vec\theta^B_0)\Big\}\Big[\tilde{\vec\theta}_n^B-\vec\theta^B_0+\Big\{\vec H_{B}(\vec\theta_0)+\vec W_n(\vec\theta^B_0)\Big\}^{-1}\vec b_n\Big].
\end{align*}
Since by the condition (C.a) $n^{-1/2}\vec S^{\text{eff}}_B(\vec\theta_0) \xrightarrow[]{d} N(\vec 0, \vec I_B(\vec\theta_0))$, we have
\begin{align*}
    &n^{1/2}\Big\{\vec H_{B}(\vec\theta_0)+\vec W_n(\vec\theta^B_0)\Big\}\Big[\tilde{\vec\theta}_n^B-\vec\theta^B_0+\Big\{\vec H_{B}(\vec\theta_0)+\vec W_n(\vec\theta^B_0)\Big\}^{-1}\vec b_n\Big] \\
    &\approx -n^{-1/2}\vec S^{\text{eff}}_B(\vec\theta_0) + \vec o_p(1) \xrightarrow[]{d} N(\vec 0, \vec I_B(\vec\theta_0)).
\end{align*}

\subsection{Computation of the sandwich variance estimate} \label{appendix.sandwich}

\noindent
For binary treatment, we assume the following logistic regression model for the pooled data
\begin{align*}
    \log\Bigg\{\frac{P(A_{ij}=1|\vec H_{ij})}{1-P(A_{ij}=1|\vec H_{ij})}\Bigg\} = \vec H_{ij}\vec\beta,
\end{align*}
where $\vec\beta$ is the vector of treatment model parameters. Usually, $\vec\beta$ is unknown and needs to be estimated using the data. Hence, for appropriate estimation of the asymptotic variance of $\tilde{\vec\psi}_n$, we need to account for the uncertainty associated with the estimation of $\vec\beta$. 
The joint score of $\hat{\vec\beta}$ and $\tilde{\vec\psi}_n$ is
\begin{align*}
    S_i &= \begin{bmatrix}
    S_{\vec\beta,(i)}\\
    S_{\vec\psi,(i)}
    \end{bmatrix} = \begin{bmatrix}
    \vec H_i^\top\{\vec A_i-E(\vec A_i|\vec H_i)\}\\
    [\{\vec A_i-E(\vec A_i|\vec H_i)\}\cdot\vec H_i]^\top\,\vec V_i^{-1}\,(\vec Y_i- \vec A_i\cdot\vec H_i\vec\psi - \vec H_i\vec\delta)
    \end{bmatrix}.
\end{align*}
Then we calculate
\begin{align*}
    \hat{I}_n(\hat{\vec\beta},\tilde{\vec\psi})&=\sum_{i=1}^n S_i\,S_i^\top\,\big|_{\vec\beta=\hat{\vec\beta},\vec\psi=\tilde{\vec\psi}}
    = \begin{bmatrix}
    \hat{I}_{\vec\beta\vec\beta}& \hat{I}_{\vec\beta\vec\psi}\\
    \hat{I}_{\vec\psi\vec\beta}& \hat{I}_{\vec\psi\vec\psi}    
    \end{bmatrix}.
\end{align*}
By Taylor's expansion and Slutskey's theorem, $I_n(\tilde{\vec\psi})$ can be consistently estimated by
\begin{align*} \label{information.psi}
  \hat{I}_n(\tilde{\vec\psi})=\hat{I}_{\vec\psi\vec\psi} -\hat{I}_{\vec\psi\vec\beta}\hat{I}_{\vec\beta\vec\beta}^{-1}\hat{I}_{\vec\beta\vec\psi}.  
\end{align*}
The estimate of $H_n(\tilde{\vec\psi})$ is
\begin{align*}
    \hat{H}_n(\tilde{\vec\psi}) &= -\sum_{i=1}^n\frac{\partial S_{\vec\psi,(i)}}{\partial\vec\psi^\top}\,\big|_{\vec\beta=\hat{\vec\beta},\vec\psi=\tilde{\vec\psi}}\\
    &=\sum_{i=1}^n[\{\vec a_i-E(\vec A_i|\vec H_i)\}\cdot\vec h_i]^\top\,\hat{\vec V}_i^{-1}\,[\{\vec a_i-E(\vec A_i|\vec H_i)\}\cdot\vec h_i]\,\big|_{\vec\beta=\hat{\vec\beta},\vec\psi=\tilde{\vec\psi}}.
\end{align*}
Note that $\vec S_{\vec\psi,(i)}=\vec S_{\vec\psi,(i)}^{\text{eff}}$.

\subsection{Additional simulation results}\label{sim.additional}

\subsubsection{Setup with unequal number of observations from patients}

We chose the number of observations of each subject from the set \{4, 5 and 6\} with an equal selection probability 1/3. The performance of our estimator did not change (compare Table~\ref{unequal.J.tab} to the bottom row of Table~\ref{large.n.tab} of the main paper).

\begin{table}[ht]
\centering
\caption{Model selection performance of the penalized G-estimator for data generated with $n=500$, unequal clusters having size 4, 5 or 6, an exchangeable correlation structure among the repeated outcomes with correlation parameter $\alpha=0.8$, an autocorrelation coefficient $\rho=0.25$ deciding the correlation among the EM's and noise covariates, and error variance $\sigma^2_\epsilon=1$. Results are obtained with a misspecified treatment-free model from 500 independent simulations for Setting 1 and Setting 2.}
\begin{tabular}{cccccccccc}
  \toprule
Working & \multicolumn{4}{c}{\bf Setting 1 (Stronger EM)}&& \multicolumn{4}{c}{\bf Setting 2 (Weaker EM)}\\
  \cline{2-5}\cline{7-10}
correlation & FN & FP & EXACT & AFP &  & FN & FP & EXACT & AFP \\ 
  \hline\hline
Indep & 0.2 & 6.0 & 93.8 & 6.00 &  & 0.6 & 5.6 & 94.0 & 6.00 \\ 
  \bf Exch & 0.4 & 6.0 & \bf 93.6 & 6.00 &  & 0.8 & 4.8 & \bf 94.6 & 4.80 \\ 
  UN & 1.4 & 5.0 & 93.6 & 5.00 &  & 1.4 & 5.2 & 93.6 & 5.40 \\  
   \bottomrule
\end{tabular} \label{unequal.J.tab}\\
\footnotesize
FN: \% of false negatives, FP: \% of false positives, EXACT: \% of exact selections, AFP: average false positives,\\ EM: effect modifier, Indep: independent, Exch: exchangeable, UN: unstructured 
\end{table}

\subsubsection{Setup where past outcome affects future exposure and future outcome}
We performed additional simulations considering a situation when past outcome is allowed to affect future treatment decision, i.e., we have effects like $Y_{j-1} \rightarrow A_j$. We generated the binary exposure according to the probability
\begin{equation}\label{supp.treat.mod}
    P(A_j = 1 | \vec H_j) = \dfrac{\exp{(\beta_0+\beta_1 l^{(1)}+\beta_2 l^{(2)}+\sum_{m=3}^6 \beta_m l_j^{(m)} + \beta_7 a_{j-1} + \beta_8 y_{j-1})}}{1+\exp{(\beta_0+\beta_1 l^{(1)}+\beta_2 l^{(2)}+\sum_{m=3}^6 \beta_m l_j^{(m)} + \beta_7 a_{j-1} + \beta_8 y_{j-1})}},
\end{equation}
where the coefficients $\beta_0,\ldots,\beta_7$ were the same as considered in other simulations presented in the main manuscript and we set $\beta_8=-0.8$. The results regarding the model-selection consistency are shown in Table~\ref{supp.tab.dag2}.

\begin{table}[ht]
\centering
\caption{Model selection performance of the proposed penalized G-estimator under a data generating mechanism when we additionally have $Y_{j-1} \rightarrow A_j$. Data were generated with $n=200$ vs.\ 500, $n_i=6$ for all $i$, an exchangeable correlation structure among the repeated outcomes with $\alpha=0.8$, an autocorrelation coeffcient $\rho=0.25$ deciding the correlation among the EM's and noise covariates, and error variance $\sigma^2_\epsilon=1$. Results are obtained with a misspecified treatment-free model from 500 independent simulations for Setting 1 and Setting 2.}
\begin{tabular}{ccccccccccc}
  \toprule
& Working & \multicolumn{4}{c}{\bf Setting 1 (Stronger EM)}&& \multicolumn{4}{c}{\bf Setting 2 (Weaker EM)}\\
  \cline{3-6}\cline{8-11}
& correlation & FN & FP & EXACT & AFP &  & FN & FP & EXACT & AFP \\ 
  \hline
 & Indep & 11.8 & 5.4 & 83.0 & 0.06 &  & 35.8 & 5.6 & 61.4 &0.07 \\ 
 $n=200$ & \bf Exch & 12.2 & 5.2 & \bf 82.8 & 0.06 &  & 36.6 & 4.4 & \bf 61.2 &0.05 \\ 
   & UN & 13.2 & 5.0 & 81.8 & 0.05 &  & 35.6 & 3.6 & 62.0 & 0.04\vspace{.2cm}\\
  & Indep & 1.4 & 4.0 & 94.6 & 0.04 &  & 1.6 & 2.6 & 96.0 & 0.03 \\ 
  $n=500$ & \bf Exch & 1.4 & 3.6 & \bf 95.0 & 0.04 &  & 1.8 & 2.2 & \bf 96.2 & 0.02 \\ 
   & UN & 1.8 & 2.8 & 95.4 & 0.03 &  & 1.8 & 2.0 & 96.2 & 0.02 \\ 
   \bottomrule
\end{tabular} \label{supp.tab.dag2}\\
\footnotesize
FN: \% of false negatives, FP: \% of false positives, EXACT: \% of exact selections,\\
AFP: average false positives, EM: effect modifier, \\
Indep: independent, Exch: exchangeable, UN: unstructured
\end{table}

We performed another simulation study where the treatment generating mechanism was the same as shown in equation (\ref{supp.treat.mod}), and additionally, we allowed past outcome to affect future outcome. In this case, the treatment-free model we considered is
\begin{align}
\mu_j(\vec h_j; \vec\delta)=\delta_0+\delta_1 l^{(1)}+\delta_2 l^{(2)}+\sum_{m=3}^6 \delta_m l_j^{(m)}+\delta_7 \exp(l_j^{(5)})+ \delta_8 a_{j-1}+ \delta_9 y_{j-1},
\end{align}
where the coefficients $\delta_0,\ldots,\delta_8$ were the same as considered in other simulations presented in the main manuscript and we set $\delta_9=0.7$.  The results regarding the model-selection consistency are shown in Table~\ref{supp.tab.dag3}. Under both setups (see Tables~\ref{supp.tab.dag2} and~\ref{supp.tab.dag3}), the selection rates of the true effect modifiers were good when we considered a larger sample size, specifically for the setting of weaker effect modification.

\begin{table}[ht]
\centering
\caption{Model selection performance of the proposed penalized G-estimator under a data generating mechanism when we additionally have $Y_{j-1} \rightarrow A_j$ and $Y_{j-1} \rightarrow Y_j$. Data were generated with $n=200$ vs.\ 500, $n_i=6$ for all $i$, an exchangeable correlation structure among the repeated outcomes with $\alpha=0.8$, an autocorrelation coeffcient $\rho=0.25$ deciding the correlation among the EM's and noise covariates, and error variance $\sigma^2_\epsilon=1$. Results are obtained with a misspecified treatment-free model from 500 independent simulations for Setting 1 and Setting 2.}
\begin{tabular}{ccccccccccc}
  \toprule
& Working & \multicolumn{4}{c}{\bf Setting 1 (Stronger EM)}&& \multicolumn{4}{c}{\bf Setting 2 (Weaker EM)}\\
  \cline{3-6}\cline{8-11}
& correlation & FN & FP & EXACT & AFP &  & FN & FP & EXACT & AFP \\ 
  \hline
  & Indep & 14.2 & 23.2 & 65.8 & 0.35 &  & 46.2 & 25.2 & 41.2 & 0.34 \\ 
 $n=200$  & \bf Exch & 13.4 & 24.4 & \bf 65.0 & 0.35 &  & 42.6 & 23.8 & \bf 45.6 & 0.33 \\ 
   & UN & 13.8 & 22.0 & 66.6 & 0.31 &  & 43.0 & 24.4 & 44.6 & 0.34\vspace{.2cm}\\
 & Indep & 2.0 & 16.4 & 81.6 & 0.20 &  & 7.8 & 15.2 & 78.0 & 0.20 \\ 
 $n=500$  & \bf Exch & 2.0 & 17.4 & \bf 80.6 & 0.22 &  & 7.4 & 16.2 & \bf 77.4 & 0.21 \\ 
   & UN & 0.8 & 17.0 & 82.2 & 0.21 &  & 6.4 & 16.2 & 78.8 & 0.20 \\ 
   \bottomrule
\end{tabular} \label{supp.tab.dag3}\\
\footnotesize
FN: \% of false negatives, FP: \% of false positives, EXACT: \% of exact selections,\\
AFP: average false positives, EM: effect modifier, \\
Indep: independent, Exch: exchangeable, UN: unstructured
\end{table}

\subsubsection{Simulations in high-dimensional setting}
To generate the data for $j$-th session ($j=1,\ldots,J$) of each subject, we generated one baseline confounder as $L^{(1)} \sim N(0,1)$, and the time varying confounders and noise covariates as $L_j^{(2)},\ldots,L_j^{(5)},X_j^{(1)},\ldots,X_j^{(K-5)} \sim MVN_{K-1}\Big((\vec\mu_{L,j}^\top,\vec\mu_{X,j}^\top)^\top,\vec V\Big)$, where $\mu_{L,j}^{(k)}= 0.3\,l^{(k)}_{j-1}+0.3\,a_{j-1}$ for $k=2,3,4 \text{ and } 5$, and $\mu_{X,j}^{(r)}= 0.5\,x^{(r)}_{j-1}$ for $ r=1,\ldots,K-5$. The covariance matrix $\vec V$ has $(r,s)$-th element equal to $\rho^{|r-s|}$ for $r,s=1,\ldots,K-1$. We generated the binary exposure according to the probability
\begin{align}
    P(A_j = 1 | \vec H_j) = \dfrac{\exp{(\beta_0+\beta_1 l^{(1)}+\sum_{m=2}^5 \beta_m l_j^{(m)} + \beta_6 a_{j-1})}}{1+\exp{(\beta_0+\beta_1 l^{(1)}+\sum_{m=2}^5 \beta_m l_j^{(m)} + \beta_6 a_{j-1})}}.
\end{align}
The generation of $\vec\epsilon$ is same as before. We constructed the outcome as $y_j = \mu_j(\vec h_j; \vec\delta) + \gamma^*_j(a_j,\vec h_j; \vec\psi) + \epsilon_j$, where
\begin{align*}
    \mu_j(\vec h_j; \vec\delta)&=\delta_0+\delta_1 l^{(1)}
    +\sum_{m=2}^5 \delta_m l_j^{(m)}
    +\delta_{6} a_{j-1}
    +\sum_{m=1}^{20} \delta_{6+m}\,x_j^{(m)}
    +\sum_{m=21}^{K-5} \delta_{6+m}\,x_j^{(m)}\\&\hspace{2cm}
    +\delta_{K+2} l^{(1)}l_j^{(4)}
    +\delta_{K+3} l_j^{(2)}l_j^{(3)}
    +\delta_{K+4} \sin(l_j^{(3)} - l_j^{(4)})
    +\delta_{K+5} \cos(2l_j^{(5)})    
\end{align*}
and 
$\gamma^*_j(a_j,\vec h_j; \vec\psi)=(\psi_0+\psi_1 l^{(1)}+\sum_{m=2}^5\psi_m l_j^{(m)}+ \psi_6 a_{j-1}+\sum_{m=1}^{20} \psi_{6+m}\,x_j^{(m)}+\sum_{m=21}^{K-5} \psi_{6+m}\,x_j^{(m)})a_j$.
Let $\vec\beta=(\beta_0,\ldots,\beta_6)^\top$, $\vec\delta=(\delta_0,\ldots,\delta_{K+5})^\top$ and $\vec\psi=(\psi_0,\ldots,\psi_{K+1})^\top$. We set
\begin{align*}
   \vec\beta&=(0,1,1,1,1,1,-0.8)^\top \\
   \vec\delta&=(1,1, 1.2, -1, 1, -1.2, 1, 1, \ldots, 1, 0, \ldots,0,-0.8, 1, 1.2, -2)^\top\\
   \vec\psi&=(1, 1.5, 1.2, -1.3, 0, 1, 2, 0, \ldots,0, 0, \ldots,0)^\top
\end{align*}
Note that $X^{(1)}$ to $X^{(20)}$ have impact on the outcome only and the coefficients of $X^{(21)}$ to $X^{(K-5)}$ were set to zero in $\mu_j(\vec h_j; \vec\delta)$. Though we set the coefficients of all the $X$'s to zero in $\gamma^*_j(a_j,\vec h_j; \vec\psi)$, it is possible to investigate effect heterogeneity by the $X$ variables also. We set $K=50$ vs.\ 100, $n=200$ vs.\ 500, $\rho=0.3$, $\sigma^2_\epsilon=1$, and $\alpha=0.8$.

When we performed penalized estimation the outcome model was misspecified, because
\begin{itemize}
    \item $L^{(1)}\times L^{(4)}$ and $L^{(2)}\times L^{(3)}$ interactions were ignored
    \item $\sin(L^{(3)}-L^{(4)})$ and $\cos(2 L^{(5)})$ terms were ignored
    \item Covariate $X^{(10)}$ which also affects the outcome was considered unmeasured
\end{itemize}
The selection performance of the estimator was poor in small samples and increasing the number of variables from 50 to 100 was associated with an increase in the false negative rates (see Table~\ref{supp.tab.high}). However, the selection performance was good when we increased the sample size from 200 to 500.

\begin{table}[ht]
\centering
\caption{Model selection performance of the proposed penalized G-estimator in high dimension. Data were generated with $n=200$ vs.\ 500, $n_i=6$ for all $i$, an exchangeable correlation structure among the repeated outcomes with $\alpha=0.8$, an autocorrelation coeffcient $\rho=0.3$ deciding the correlation among the EM's and noise covariates, and error variance $\sigma^2_\epsilon=1$. Results are obtained with a misspecified treatment-free model from 500 independent simulations for two different values of dimension $K$.}
\begin{tabular}{ccccccccccc}
  \toprule
& Working & \multicolumn{4}{c}{}&& \multicolumn{4}{c}{}\\
& correlation & FN & FP & EXACT & AFP &  & FN & FP & EXACT & AFP \\ 
  \hline
  & & \multicolumn{4}{c}{$\mathbf{n=200}$}&& \multicolumn{4}{c}{$\mathbf{n=500}$}\\
  \cline{3-6}\cline{8-11}
 & Indep & 34.0 & 1.2 & 64.8 & 1.20 &  & 2.4 & 2.0 & 95.6 & 2.00 \\ 
 $\mathbf{K=50}$  & \bf Exch & 35.4 & 1.2 & \bf 63.4 & 1.20 &  & 2.6 & 1.6 & \bf 95.8 & 1.60 \\ 
   & UN & 35.0 & 1.4 & 63.6 & 1.40 &  & 2.6 & 1.8 & 95.6 & 1.80\vspace{0.2cm}\\ 
   & Indep & 44.4 & 0.0 & 55.6 & 0.00 &  & 6.2 & 0.4 & 93.4 & 0.40 \\ 
 $\mathbf{K=100}$  & \bf Exch & 48.2 & 0.2 & \bf 51.6 & 0.20 &  & 6.4 & 0.0 & \bf 93.6 & 0.00 \\ 
   & UN & 47.2 & 0.2 & 52.6 & 0.20 &  & 6.6 & 0.0 & 93.4 & 0.00 \\ 
   \bottomrule
\end{tabular} \label{supp.tab.high}\\
\footnotesize
FN: \% of false negatives, FP: \% of false positives, EXACT: \% of exact selections,\\
AFP: average false positives, EM: effect modifier, \\
Indep: independent, Exch: exchangeable, UN: unstructured
\end{table}

\clearpage
\subsubsection{Comparison with the method in \cite{boruvka2018assessing}}

\begin{table}[ht]
\centering
\caption{Performance comparison between our penalized estimates (penalized-G) and the full model estimates obtained using the method in Boruvka (2018). Estimation was performed with data generated using $n=200$ and 500, autocorrelation coefficient $\rho = 0.25$, error variance $\sigma_\epsilon^2 = 1$, $\alpha=0.8$  and an exchangeable correlation structure. Statistics were calculated from 500 simulations under the independent working correlation structure
.}
\resizebox{\textwidth}{!}{
\begin{tabular}{lcccccccccccccccc}
  \toprule
 & &\multicolumn{7}{c}{$\mathbf{n=200}$}&&\multicolumn{7}{c}{$\mathbf{n=500}$}\\
  \cline{3-9}\cline{11-17}
 &True &\multicolumn{3}{c}{\bf Penalized-G}&&\multicolumn{3}{c}{\bf Boruvka (2018)}&&\multicolumn{3}{c}{\bf Penalized-G}
 & &\multicolumn{3}{c}{\bf Boruvka (2018)}\\
  \cline{3-5}\cline{7-9}\cline{11-13}\cline{14-17}
 & Coef & Bias$^*$ & SE1 & SE2 & & Bias & SE1 & SE2 & & Bias & SE1 & SE2 & & Bias & SE1 & SE2 \\ 
  \hline
$A$ & 1 & 0.20 & 0.27 & 0.24 &  & 0.74 & 0.31 & 0.51 &  & 0.30 & 0.16 & 0.16 &  & 0.64 & 0.19 & 0.31 \\ 
  $A\times L^{(1)}$& -2.5 & 0.29 & 0.32 & 0.30 &  & 1.52 & 0.40 & 0.61 &  & 0.58 & 0.19 & 0.19 &  & 0.15 & 0.24 & 0.37 \\ 
  $A\times L^{(2)}$& 1.5 & 0.03 & 0.18 & 0.17 &  & 0.07 & 0.23 & 0.32 &  & 0.02 & 0.11 & 0.11 &  & 0.17 & 0.15 & 0.19 \\ 
  $A\times L^{(3)}$& 1.5 & 0.03 & 0.18 & 0.16 &  & 0.05 & 0.24 & 0.31 &  & 0.00 & 0.11 & 0.10 &  & 0.14 & 0.14 & 0.18 \\ 
  $A\times L^{(4)}$& 1.5 & 0.03 & 0.18 & 0.17 &  & 0.16 & 0.24 & 0.32 &  & 0.13 & 0.11 & 0.11 &  & 0.06 & 0.14 & 0.19 \\ 
  $A\times L^{(5)}$& 1.5 & 0.16 & 0.31 & 0.25 &  & 0.82 & 0.35 & 0.42 &  & 0.02 & 0.18 & 0.18 &  & 0.14 & 0.21 & 0.26 \\ 
  $A\times L^{(6)}$& 0 & 0.22 & 0.07 & 0.03 &  & 0.17 & 0.24 & 0.35 &  & 0.22 & 0.05 & 0.02 &  & 0.46 & 0.14 & 0.21 \\ 
  $A\times A_{\text{Lag1}}$& 2 & 0.01 & 0.38 & 0.31 &  & 1.21 & 0.41 & 0.68 &  & 0.09 & 0.19 & 0.20 &  & 0.76 & 0.23 & 0.41 \\ 
  $A\times X^{(1)}$& 0 & 0.02 & 0.03 & 0.01 &  & 0.21 & 0.21 & 0.21 &  & 0.02 & 0.02 & 0.01 &  & 0.21 & 0.13 & 0.13 \\ 
  $A\times X^{(2)}$& 0 & 0.01 & 0.01 & 0.01 &  & 0.01 & 0.20 & 0.21 &  & 0.01 & 0.01 & 0.00 &  & 0.02 & 0.13 & 0.13 \\ 
  $A\times X^{(3)}$& 0 & 0.01 & 0.02 & 0.01 &  & 0.28 & 0.21 & 0.21 &  & 0.00 & 0.00 & 0.00 &  & 0.07 & 0.12 & 0.12 \\ 
  $A\times X^{(4)}$& 0 & 0.00 & 0.00 & 0.00 &  & 0.09 & 0.20 & 0.21 &  & 0.00 & 0.00 & 0.00 &  & 0.05 & 0.12 & 0.12 \\ 
  $A\times X^{(5)}$& 0 & 0.01 & 0.02 & 0.01 &  & 0.05 & 0.20 & 0.21 &  & 0.01 & 0.01 & 0.00 &  & 0.23 & 0.11 & 0.12 \\ 
  $A\times X^{(6)}$& 0 & 0.00 & 0.00 & 0.00 &  & 0.10 & 0.20 & 0.21 &  & 0.02 & 0.03 & 0.01 &  & 0.15 & 0.12 & 0.12 \\ 
  $A\times X^{(7)}$& 0 & 0.00 & 0.00 & 0.00 &  & 0.27 & 0.21 & 0.21 &  & 0.01 & 0.01 & 0.01 &  & 0.14 & 0.12 & 0.12 \\ 
  $A\times X^{(8)}$& 0 & 0.01 & 0.03 & 0.01 &  & 0.08 & 0.21 & 0.21 &  & 0.00 & 0.00 & 0.00 &  & 0.05 & 0.12 & 0.12 \\ 
  $A\times X^{(9)}$& 0 & 0.01 & 0.02 & 0.01 &  & 0.08 & 0.22 & 0.21 &  & 0.00 & 0.00 & 0.00 &  & 0.11 & 0.12 & 0.12 \\ 
  $A\times X^{(10)}$& 0 & 0.01 & 0.02 & 0.01 &  & 0.16 & 0.19 & 0.20 &  & 0.00 & 0.00 & 0.00 &  & 0.08 & 0.12 & 0.12 \\ 
   \bottomrule
\end{tabular}\label{supp.tab.boruvka}}
\footnotesize
 $^*$Bias (scaled bias) $=\sqrt{n}\times |\widetilde{\psi}_n-\psi_0|$, SE1: the empirical root-mean squared error (MSE), SE2: the square-root of the average of sandwich variance estimates
\end{table}

\noindent
Our estimator has higher efficiency than Boruvka's estimator, because our method applies simultaneous elimination of spurious effect modifiers.

\subsection{Descriptive statistics for hemodiafiltration data} \label{hdf.descriptives}

\begin{figure}[ht]   
    \centering
    \includegraphics[scale=0.9]{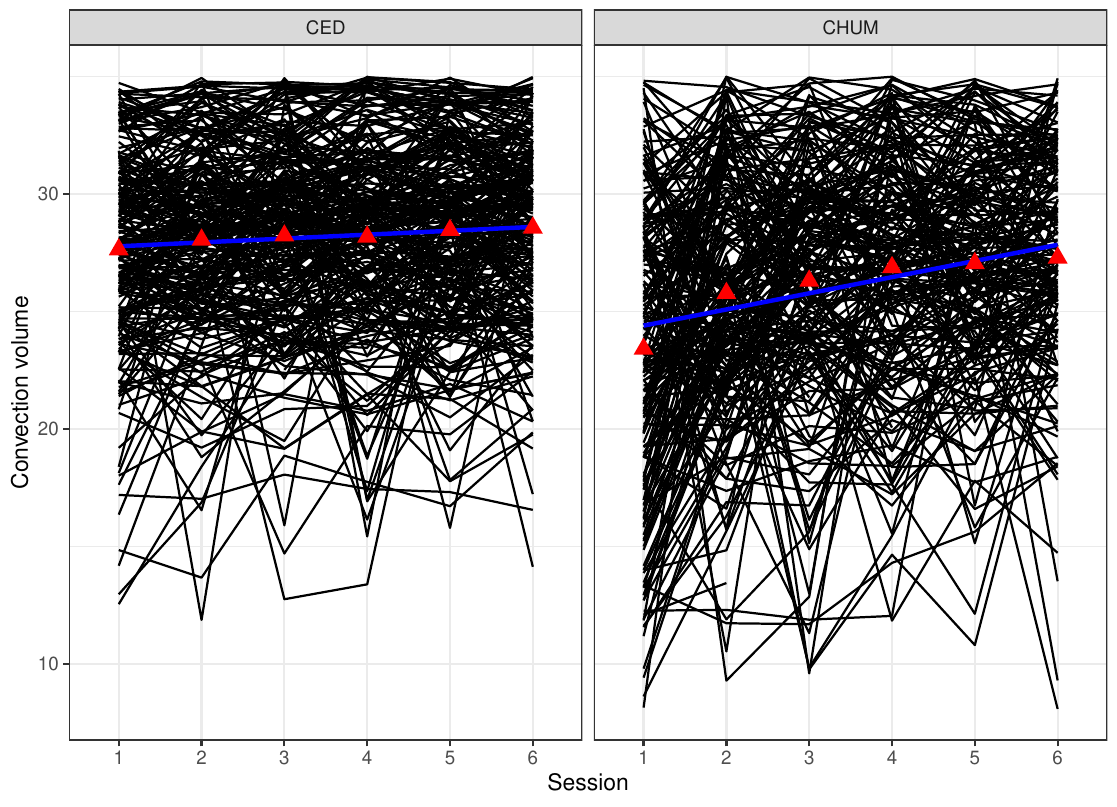}
    \caption{Individual session-specific outcomes (convection volume) of all patients by exposure status (dialysis facility). Red triangles show the session-specific means and the blue lines show the linear fits.}
    \label{fig.outcome.by.exposure}
\end{figure}

Descriptive statistics for all of the variables (continuous and binary) from Session 1 are presented in Table~\ref{tab:descriptive} by the two exposure categories (CED vs.\ CHUM). For continuous variables, we presented the mean and standard deviation (SD), and for binary variables we presented the count ($n$) and percentage. Session-specific mean outcomes of all patients from CED and CHUM are presented in Figure~\ref{fig.outcome.by.exposure}. The number of patients who switched their dialysis locations are presented in Table~\ref{tab:switch} by different sessions.

\begin{table}[ht]
\centering
\caption{Descriptive statistics of data from Session 1.}
\begin{tabular}{lrrr}
  \toprule
  &\multicolumn{3}{c}{\bf Dialysis facility}\\
  \cline{2-4}
\bf Variables &\ CED ($n = 237$) &  & CHUM ($n = 220$) \\ 
  \hline
\bf Continuous: Mean (SD)\\
Outcome & 27.64 (4.26) &  & 23.41 (6.28) \\ 
  Hemoglobin & 107.31 (14.59) &  & 93.95 (15.78) \\ 
  Albumin & 37.1 (3.75) &  & 33.03 (5.19) \\ 
  Dalteparin$^*$ & 47.28 (25.03) &  & 27.36 (21.03) \\
  Age & 65.64 (15.02) &  & 68.22 (13.88) \vspace{0.2cm}\\
\bf Binary: n (\%)\\
  Access catheter & 104 (0.44) &  & 171 (0.78) \\ 
  Alteplase & 7 (0.03) &  & 4 (0.02) \\ 
  Catheter change & 1 (0.00) &  & 9 (0.04) \\ 
  Male & 142 (0.60) &  & 141 (0.64) \\ 
  Hypertension & 59 (0.25) &  & 129 (0.59) \\ 
  Diabetes & 40 (0.17) &  & 93 (0.42) \\ 
  Peripheral vascular disease (pvd) & 11 (0.05) &  & 48 (0.22) \\ 
  Congestive heart failure (chf) & 13 (0.05) &  & 45 (0.20) \\ 
  Cardiac arrhythmia & 14 (0.06) &  & 36 (0.16) \\ 
  Acute myocardial infarction (ami) & 11 (0.05) &  & 35 (0.16) \\ 
  Chronic pulmonary disease (copd) & 9 (0.04) &  & 33 (0.15) \\ 
  Liver disease & 11 (0.05) &  & 33 (0.15) \\ 
  Valvular disease & 7 (0.03) &  & 16 (0.07) \\ 
  Cancer & 7 (0.03) &  & 14 (0.06) \\ 
  Metastatic cancer & 1 (0.00) &  & 1 (0.00) \\ 
  Cerebrovascular disease (cvd) & 2 (0.01) &  & 15 (0.07) \\ 
  Dementia & 2 (0.01) &  & 5 (0.02) \\ 
  Hemiplegia & 1 (0.00) &  & 10 (0.05) \\ 
  Rheumatic disease & 4 (0.02) &  & 2 (0.01) \\ 
   \bottomrule
\end{tabular}\label{tab:descriptive}
\end{table}

\begin{table}[ht]
    \centering
    \caption{Number of patients who switched their locations.}
    \begin{tabular}{ccc}
\toprule
&CHUM to CED&CED to CHUM\\
\hline
Session 2 & 5 & 1\\
Session 3 & 6 & 0\\
Session 4 & 3 & 0\\
Session 5 & 6 & 0\\
Session 6 & 3 & 3\\
\bottomrule
    \end{tabular}
    \label{tab:switch}
\end{table}

\end{document}